\documentclass[12pt,pre,aps,showpacs,preprint]{revtex4}

\usepackage{graphics}
\usepackage{float}
\usepackage{graphicx}
\usepackage{amsfonts}
\usepackage{textcomp}
\usepackage{amssymb}
\usepackage{amsmath}
\usepackage{ulem}
\usepackage{color}

\begin{document}

\title{Dense Packings of Polyhedra: Platonic and Archimedean Solids}

\author{S. Torquato$^{1,2,3,4,5,6}$ and Y. Jiao$^6$}


\affiliation{$^1$Department of Chemistry, Princeton University,
Princeton New Jersey 08544, USA}

\affiliation{$^2$Princeton Center for Theoretical Science,
Princeton University, Princeton New Jersey 08544, USA}

\affiliation{$^3$Princeton Institute for the Science and Technology of
Materials, Princeton University, Princeton New Jersey 08544, USA}

\affiliation{$^4$Program in Applied and Computational Mathematics,
Princeton University, Princeton New Jersey 08544, USA}

\affiliation{$^5$School of Natural Sciences, Institute for
Advanced Study, Princeton NJ 08540}

\affiliation{$^6$Department of Mechanical and Aerospace
Engineering, Princeton University, Princeton New Jersey 08544,
USA}

\begin{abstract}
Understanding the nature of dense particle packings is a subject of intense research 
in the physical, mathematical and biological sciences.  
The preponderance of previous work has focused on spherical particles,
and very little is known about dense polyhedral packings.
We formulate the problem of generating dense packings of 
nonoverlapping, non-tiling polyhedra  within an adaptive fundamental cell  
subject to periodic boundary conditions as  an optimization problem,
which we call the Adaptive Shrinking Cell (ASC) scheme.
 This novel optimization problem is solved 
here (using a variety of multi-particle initial configurations) 
to find the dense packings of each of the Platonic solids in
three-dimensional Euclidean space $\mathbb{R}^3$,
except for the cube, which is the only Platonic solid that tiles space.
We find the densest known packings of tetrahedra, icosahedra,
dodecahedra, and octahedra with densities {$0.823\ldots$}, 
$0.836\ldots$, $0.904\ldots$, and $0.947\ldots$, respectively. It
is noteworthy that the densest tetrahedral packing possesses
{\it no long-range order}. Unlike the densest tetrahedral packing,
which must not be a Bravais lattice packing, the densest packings
of the other non-tiling Platonic solids that we obtain are their 
previously known optimal (Bravais) lattice packings.
We also derive a simple upper bound on the maximal density of packings
of congruent nonspherical particles, and apply it to 
Platonic solids, Archimedean solids, superballs and ellipsoids.
Provided that what we term the  ``asphericity" (ratio of the circumradius to inradius) 
is sufficiently small, the upper bounds are relatively tight
and thus close to the corresponding densities
of the optimal lattice packings of the centrally symmetric Platonic and 
Archimedean solids. Our simulation results, rigorous upper bounds, and other theoretical arguments
lead us to the  conjecture that the densest packings of Platonic and Archimedean 
solids with central symmetry are given by their corresponding densest lattice packings. 
This can be regarded to be the analog of Kepler's sphere conjecture for these solids. 
 The truncated tetrahedron is the only non-centrally symmetric  Archimedean solid,
the densest known packing of which is a non-lattice
packing with density at least as high as $23/24 = 0.958333\ldots$.
We discuss the validity of our conjecture to packings of superballs, prisms 
and anti-prisms as well as to high-dimensional  analogs of the Platonic solids.
In addition, we conjecture that the optimal packing of any convex, congruent polyhedron
without central symmetry generally is not a lattice packing. 
Finally, we discuss the possible applications and generalizations of the ASC scheme in 
the predicting the crystal structures of polyhedral nanoparticles and 
the study of random packings of hard polyhedra.
\end{abstract}

\maketitle

\section{Introduction}

Particle packing problems are ancient, dating back to the
dawn of civilization. Bernal has remarked that ``heaps'' (particle packings) were the first things that were
ever measured in the form of basketfuls of grain for the purpose of trading or of collection
of taxes \cite{Ber65}. Dense packings of hard particles
have served as useful models to understand the structure of
liquid, glassy and crystal states of matter \cite{Za83,Ch00,To02}, granular media \cite{Ed94},
and heterogeneous materials \cite{To02,Ki91}. Understanding the symmetries and other mathematical properties
of the densest packings in arbitrary dimensions is a problem of long-standing interest in
discrete geometry and number theory \cite{Co98}.

A large collection of nonoverlapping solid objects (particles) in $d$-dimensional
Euclidean space $\mathbb{R}^d$ is called a packing. The packing density
$\phi$ is defined as the fraction of space $\mathbb{R}^d$ covered by the particles.
A problem that has been a source of fascination to mathematicians
and scientists for centuries is the determination of the densest
arrangement(s) of particles that do not tile space and the associated maximal density $\phi_{max}$  \cite{Co98}.
Finding the maximal-density packing arrangements  is directly relevant to
understanding the structure and
properties of crystalline equilibrium phases of particle systems
as well as their (zero-temperature) ground-state structures in low dimensions
in which the interactions are characterized by steep
repulsions and  short-ranged attractions. 

The preponderance of previous investigations have focused on dense packings
of spheres in various dimensions \cite{Co98,To02,Co03,Pa06}.  For congruent particles 
in three dimensions, the sphere
is the only non-tiling particle for which the densest packing arrangements
can be proved \cite{Ha05}. It is only very recently that attention
has turned to finding the maximal-density packing arrangements
of nonspherical particles in $\mathbb{R}^3$, including ellipsoids \cite{Wi91,Do04}, 
tetrahedra \cite{Co06,Ch07,Ch08}, 
and  superballs \cite{Ji08, Ji09}.  Very little is known about the densest packings of
polyhedral particles \cite{supra}.

\begin{figure}[bthp]
\begin{center}
\includegraphics[width=14.5cm,keepaspectratio]{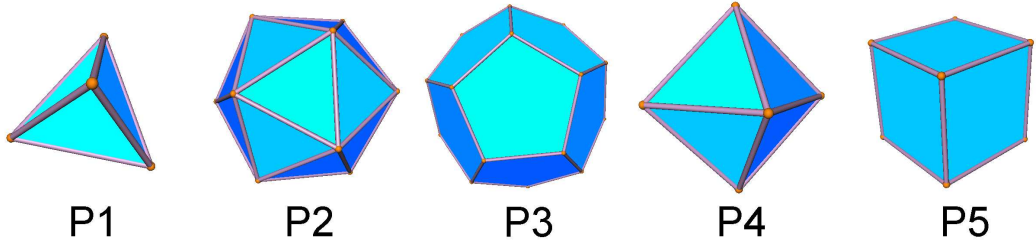} \\
\end{center}
\caption{(color online). The five Platonic solids: tetrahedron (P1), icosahedron (P2), 
dodecahedron (P3), octahedron (P4), and cube (P5).}
\label{platonic}
\end{figure}

The Platonic solids  (mentioned in Plato's Timaeus) 
are convex polyhedra with faces composed of congruent 
convex regular polygons. There are exactly five such solids: the tetrahedron (P1), icosahedron (P2), 
dodecahedron (P3), octahedron (P4), and cube (P5) (see Fig. \ref{platonic}) \cite{virus}. 
One major concern in this paper
is the determination of the densest packings of each 
of the Platonic solids in three-dimensional Euclidean space $\mathbb{R}^3$,
except for the cube, which is the only Platonic solid that tiles space.

\begin{figure}[bthp]
\begin{center}
\includegraphics[width=6cm,keepaspectratio]{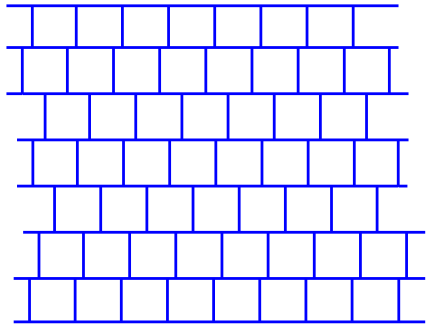} 
\end{center}
\caption{(color online). A portion of a realization of an irregular tiling
of the plane by squares.} 
\label{random-squares}
\end{figure}

It is useful to highlight some basic geometrical properties
of the Platonic solids that we will employ in subsequent sections of the
paper. The dihedral angle $\theta$ is the interior angle between any two face planes
and is given by
\begin{equation}
\sin{\theta\over 2} = \frac{\cos(\pi/q)}{\sin(\pi/p)},
\end{equation}
where $p$ is the number of sides of each face and $q$ is the number of faces meeting at each vertex.
Thus, $\theta$ is $2\sin^{-1}(1/\sqrt{3})$, $2\sin^{-1}(\Phi/\sqrt{3})$, $2\sin^{-1}(\Phi/\sqrt{\Phi^2+1})$, 
$2\sin^{-1}(\sqrt{2/3})$, and $\pi/2$, for the tetrahedron, icosahedron, dodecahedron,
octahedron, and cube, respectively, where $\Phi=(1+\sqrt{5})/2$ is the golden ratio.
Thus, since the dihedral angle for the cube is the only one
that is a submultiple of $2\pi$, the cube is the only Platonic solid
that tiles space. We note in passing that in addition
to the regular tessellation of space by cubes in the simple cubic lattice
arrangement, there are an infinite
number of other irregular tessellations of space by cubes \cite{degen}. Figure \ref{random-squares}
shows a portion of a realization of a two-dimensional analog of 
such an irregular tessellation.

Every polyhedron has a dual polyhedron with faces and vertices interchanged. The dual of each
Platonic solid is another Platonic solid, and therefore 
they can be arranged into dual pairs: the tetrahedron is self-dual (i.e., its dual is another 
tetrahedron), the icosahedron and 
dodecahedron form a dual pair, and the octahedron and cube form a dual pair.

\begin{figure}
\begin{center}
\includegraphics[width=14.5cm,keepaspectratio]{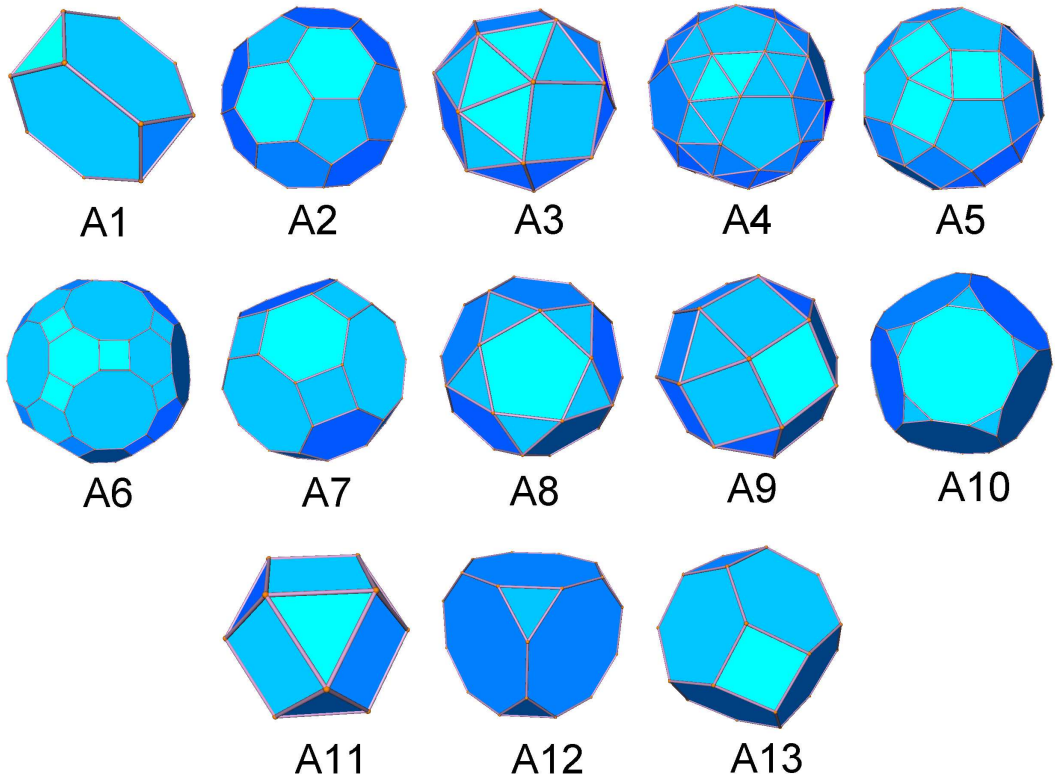} \\
\end{center}
\caption{The 13 Archimedean solids: truncated tetrahedron (A1), 
truncated icosahedron (A2),
snub cube (A3), snub dodecahedron (A4), rhombicosidodecahdron (A5), 
truncated icosidodecahdron (A6), 
truncated cuboctahedron (A7), icosidodecahedron (A8),  rhombicuboctahedron (A9),
truncated dodecahedron (A10), cuboctahedron (A11), 
truncated cube (A12), and truncated octahedron (A13).} \label{archimedean}
\end{figure}

An Archimedean solid is a highly symmetric, semi-regular
convex polyhedron composed of two or more types of regular polygons meeting in 
identical vertices. There are thirteen Archimedean solids: truncated tetrahedron (A1), 
truncated icosahedron (A2), snub cube (A3), snub dodecahedron (A4), rhombicosidodecahdron (A5), 
truncated icosidodecahdron (A6), truncated cuboctahedron (A7), icosidodecahedron (A8),  
rhombicuboctahedron (A9), truncated dodecahedron (A10), cuboctahedron (A11), 
truncated cube (A12) and truncated octahedron (A13) (see Fig. \ref{archimedean}). 
Note that the truncated octahedron is the only Archimedean solid that tiles space.

Another important observation is that the tetrahedron (P1) and 
the truncated tetrahedron (A1) are the 
only Platonic and Archimedean solids, respectively, 
that are not {\it centrally symmetric}.
A particle is centrally symmetric if it has a center $C$ that
bisects every chord through $C$ connecting any two boundary points
of the particle, i.e., the center is a point of inversion symmetry.
 We will see that the central symmetry
of the majority of the Platonic and Archimedean solids (P2 -- P5, A2 -- A13) 
distinguish their dense packing arrangements from those of the
non-centrally symmetric ones (P1 and A1) 
in a fundamental way.

Some basic definitions concerning packings are given here. A {\it saturated} packing 
is one in which there is no space available to add another particle to the packing.
A {\it lattice} $\Lambda$ in $\mathbb{R}^d$ is a subgroup
consisting of the integer linear combinations of vectors that constitute 
a basis for $\mathbb{R}^d$ \cite{lattice}.
A {\it lattice packing} $P_L$ is one in which  the centroids of the nonoverlapping particles
are located at the points of $\Lambda$, each oriented in the same direction.
The set of lattice packings is a subset of all possible packings in $\mathbb{R}^d$.
In a lattice packing, the space $\mathbb{R}^d$ can be geometrically divided into identical
regions $F$ called {\it fundamental cells}, each of which contains 
just the centroid of one particle. Thus, the density of a lattice packing is given by
\begin{equation}
\phi= \frac{v_p}{\mbox{Vol}(F)},
\end{equation}
where $v_p$ is the volume of a $d$-dimensional particle and
$\mbox{Vol}(F)$ is the volume of a fundamental cell.

A more general notion than a lattice packing is a periodic
packing. A {\it periodic} packing of congruent particles
is obtained by placing a fixed nonoverlapping configuration of $N$ particles (where $N\ge 1$)
with {\it arbitrary orientations} in each fundamental cell of a lattice $\Lambda$. Thus, the packing is still
periodic under translations by $\Lambda$, but the $N$ particles can occur
anywhere in the chosen fundamental cell subject to the nonoverlap condition.
The packing density of a  periodic packing is given by
\begin{equation}
\phi=\frac{N v_p}{\mbox{Vol}(F)}=\rho v_p,
\end{equation}
where $\rho=N/\mbox{Vol}(F)$ is the number density, i.e., the number of particles per unit volume.

The determination of the maximal-density arrangements of non-tiling
polyhedral particles is a notoriously difficult problem, especially
since such extremal structures will generally be non-Bravais-lattice
packings. Computer simulations that seek the maximal-density packings
can be an indispensable tool, especially if they can incorporate
collective motions of the particles in order to obtain,
in principle,  the highest possible densities. However, the challenge presented by polyhedral
particles in $\mathbb{R}^3$ is the non-smooth (i.e., nonanalytic)
nature of the particle shape. In the case of smoothly-shaped particles,
such as spheres, ellipsoids and superballs, one can construct analytic
``overlap potential functions" for the particles \cite{overlap} and hence
one can employ efficient collision-driven molecular dynamics (MD) growth packing algorithms that inherently
involve collective particle motions \cite{Do04,Ji08,Ji09,Do05a,Do05b}.
The fact that analytic overlap potential functions
cannot be constructed for polyhedral particles prevents us
from using event-driven MD growth methods
to study such systems.

In this paper, we devise a novel optimization scheme,
called the Adaptive Shrinking Cell (ASC), 
that can be applied to generate dense packings of polyhedra in $\mathbb{R}^3$.
We employ it specifically to obtain the densest known packings of tetrahedra, icosahedra,
dodecahedra, and octahedra with densities {$0.823\dots$}, $0.836\ldots$, $0.904\ldots$,
and $0.947\ldots$, respectively. The result for tetrahedra improves
upon the density reported in our recent investigation \cite{To09a,footnote}.
Unlike the densest tetrahedral packing,
which  must not be a Bravais lattice packing \cite{Co06}, the densest packings
of the other non-tiling Platonic solids that we obtain
are their corresponding densest lattice packings \cite{Mi04,Be00}.

We also derive a simple upper bound on the maximal density of packings
of congruent nonspherical particles, and apply it to 
Platonic solids, Archimedean solids, superballs and ellipsoids. 
We introduce the  ``asphericity" 
parameter $\gamma$ (ratio of the circumradius to inradius) to show that when $\gamma$  
is sufficiently small, the upper bounds are relatively tight
and thus close to the corresponding densities
of the optimal lattice packings of octahedra, dodecahedra and icosahedra
as well as of the majority of the Archimedean solids with central symmetry.

Our simulation results as well as theoretical arguments lead us to conjecture 
that the densest packings of Platonic and Archimedean solids with central symmetry 
are given by their corresponding densest lattice packings. 
This can be regarded to be the analog of Kepler's sphere conjecture
for these solids.
The truncated tetrahedron is the only non-centrally symmetric  Archimedean solid,
the densest known packing of which is a non-lattice
packing with density as high as $23/24 = 0.958333\ldots$.
Our work also suggests that the optimal packings 
of superballs are their corresponding densest lattice packings.

In a recent letter \cite{To09a}, we briefly 
reported the densest known packings of the non-tiling Platonic solids 
obtained using the ASC algorithm and proposed the aforementioned conjecture concerning 
the optimal packings of the Platonic and Archimedean solids.
In this paper, we expand on theoretical and
computational details and report additional new results. 
In particular,  we provide comprehensive details about the ASC scheme and the 
simulation results (Sec. II), including a discussion about the
various initial configurations we used for the ASC algorithm (Sec. III). Moreover, 
we have improved on the highest tetrahedral packing density reported in Ref.~\cite{To09a} 
(i.e., from $0.782\ldots$ to $0.823\ldots$),  
by exploring a broad range of dynamical parameters for the algorithm and initial configurations.
Certain pair  statistics of the densest known packing of tetrahedra (e.g., the contact number, the centroidal 
correlation function and the face-normal correlation function) are given (Sec. III).
It is noteworthy that the densest tetrahedral packing is a non-Bravais
structure with a complex periodic cell and possesses 
{\it no long-range order}.

The initial configurations for the icosahedral, dodecahedral and octahedral 
packings are described and  the numerical challenges in producing 
dense packings of such polyhedral particles are discussed (Sec. III).
In addition, a detailed derivation of the upper bound and tables containing 
the geometrical characteristics of the Platonic and Archimedean solids 
as well as their upper bound values are given (Sec. IV). 
The upper bound is also applied to superballs and ellipsoids (which was not done
in Ref. \cite{To09a}).  
Moreover, we provide the major elements of a possible proof of 
our conjecture (Sec. V).  We also discuss how our conjecture could be generalized to 
other centrally symmetric polyhedral particles such as prisms 
and anti-prisms as well as high-dimensional 
analogs of the Platonic solids. Our work
also naturally leads to another conjecture reported
here for the first time, namely, the optimal packing of any convex, congruent polyhedron
without central symmetry is generally not a (Bravais) lattice packing (Sec. V). 

Furthermore, we discuss the possible applications and generalizations of the ASC scheme to 
predict the crystal structures of polyhedral nanoparticles and 
to the study of random packings of hard polyhedra.
Finally, we collect in appendices basic packing characteristics
of various optimal lattice and non-lattice packings of polyhedra (including lattice vectors)
that have been scattered throughout the literature and provide
lattice vectors and other characteristics of the densest known 
packings of tetrahedra (obtained here) and truncated tetrahedra.

\section{Adaptive Shrinking Cell (ASC) Optimization Scheme}

We formulate the problem of generating dense packings of 
nonoverlapping, non-tiling polyhedra  within a fundamental cell in $\mathbb{R}^3$ 
subject to periodic boundary conditions as an optimization problem.
In particular, the objective function
is taken to be the negative of the packing density $\phi$. Starting from an initial unsaturated packing
configuration of particles of fixed size in the fundamental cell, the positions and orientations 
of the polyhedra are design variables for the optimization.
Importantly, we also allow the boundary of the fundamental cell
to deform macroscopically as well as compress or expand (while
keeping the particles fixed in size) such that there is a net compression
(increase of the density of the packing) in the final state \cite{Deform}. Thus,
the deformation and compression/expansion  of the cell
boundary, which we call the {\it adaptive fundamental cell},
are also design variables. We are not aware of any packing algorithm
that employs both a {\it sequential} search of the configurational space of 
the particles and the space of lattices via an adaptive fundamental cell that shrinks
on average to obtain dense packings. We will call this optimization scheme the Adaptive
Shrinking Cell (ASC).

\begin{figure}[bthp]
\begin{center}
$\begin{array}{c@{\hspace{0.5cm}}c@{\hspace{0.5cm}}c}\\
\includegraphics[width=5.0cm,keepaspectratio]{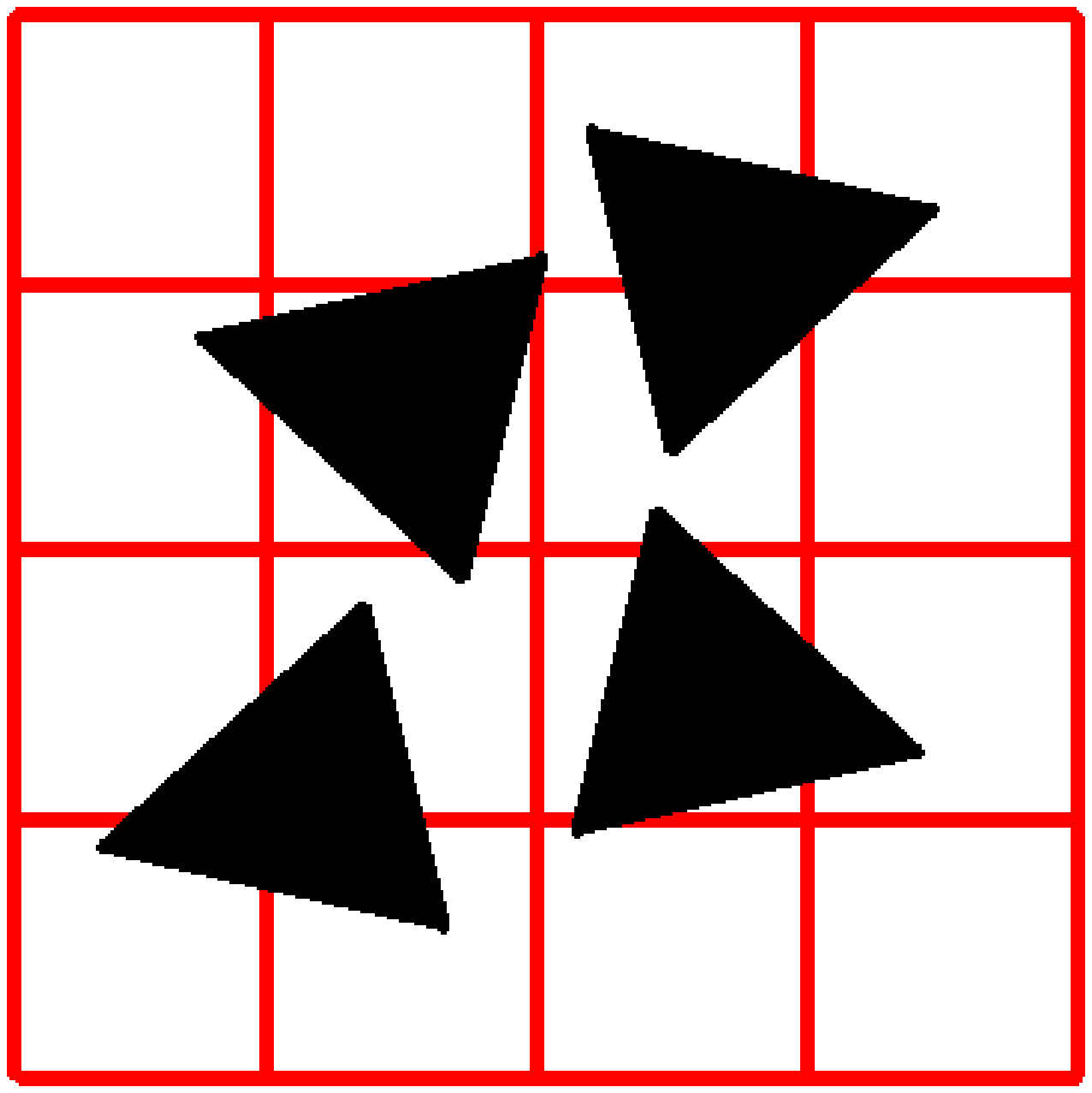} &
\includegraphics[width=5.0cm,keepaspectratio]{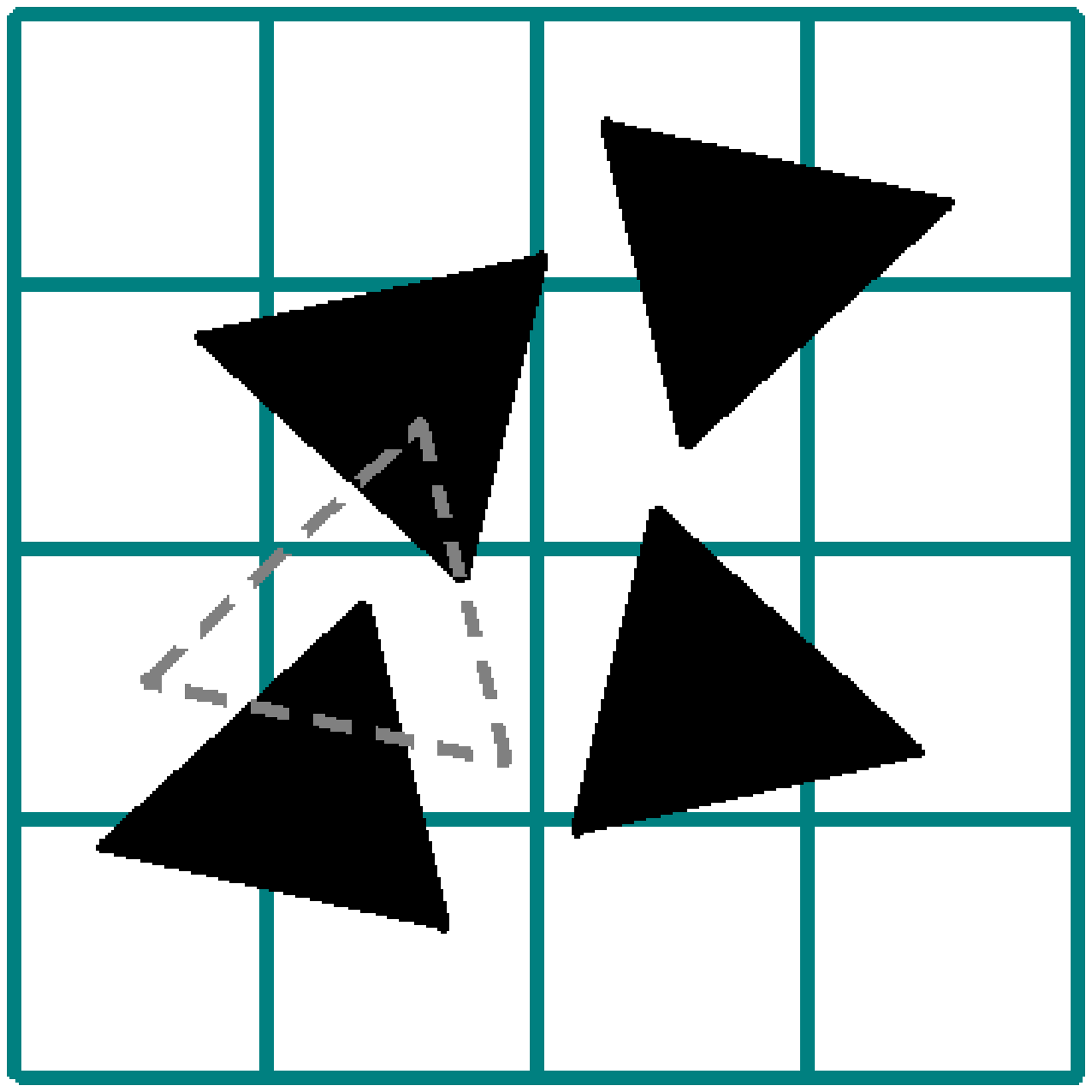} &
\includegraphics[width=5.0cm,keepaspectratio]{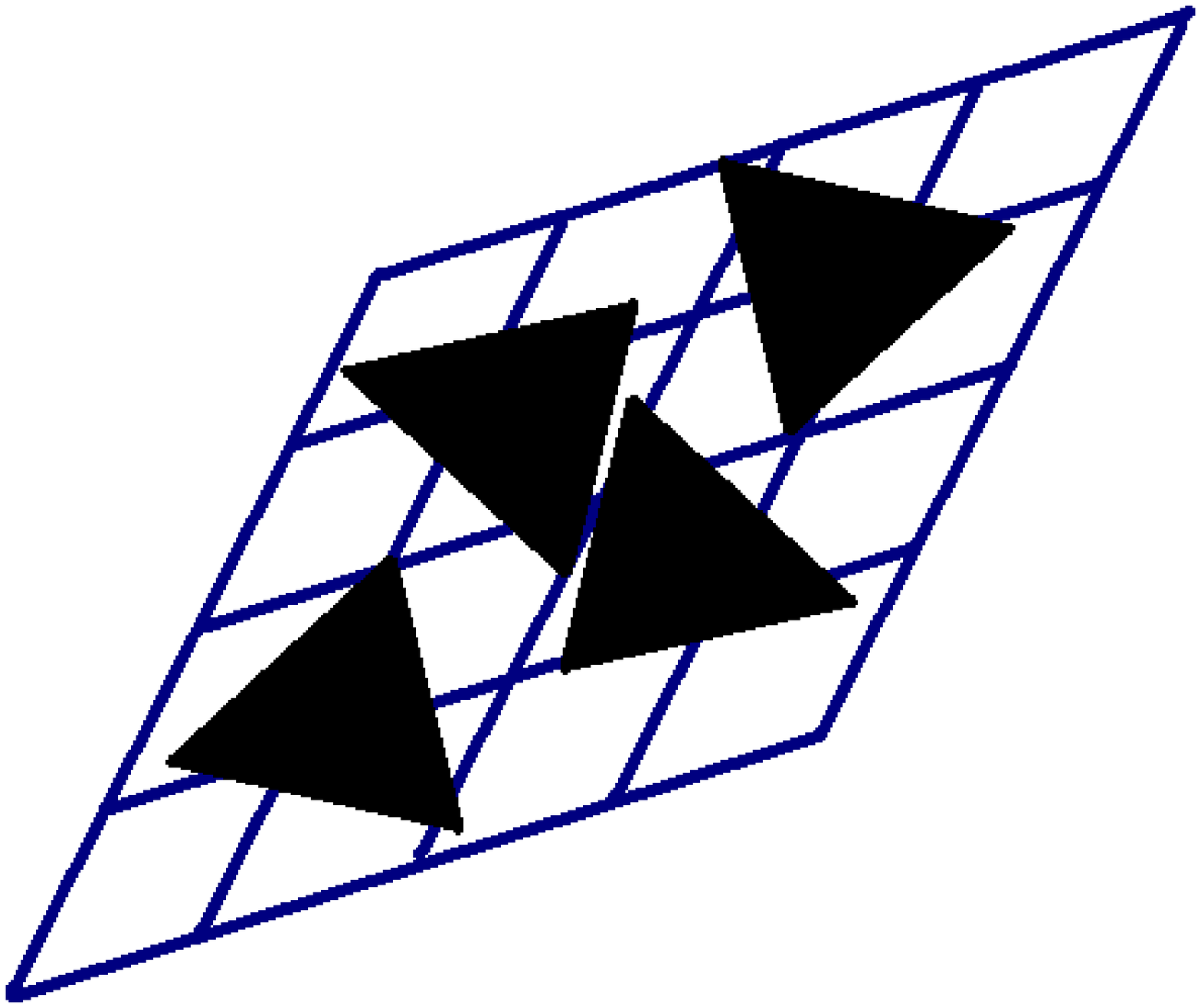} \\
\mbox{\bf (a)} & \mbox{\bf (b)} & \mbox{\bf (c)}
\end{array}$
\end{center}
\caption{(color online). Sequential changes of the packing configuration
due to the design variables in the ASC algorithm. (a) An initial configuration of four particles. (b) A trial
move of a randomly selected particle that is rejected because it overlaps another particle.
(c) A trial move that is accepted, which results
in a deformation and compression (small in magnitude) changing
the fundamental cell shape and size as well as the relative distances between the particles. The large 
fundamental cell is divided into smaller sub-cells in order to implement the 
``cell method'' discussed in Sec.~II.D.}
\label{fig_Cell}
\end{figure}

We will see that the  ASC optimization scheme allows for some desired collective 
motions of the particles to find the optimal lattice for the periodic cell. This is
to be contrasted with previous treatments that use  a fixed shape
for the fundamental cell, which may or may not be the optimal shape.   
Figure \ref{fig_Cell} illustrates a simple sequence of configuration
changes for a four-particle system (which is explained in more detail
in Sec. II.A).
By efficiently exploring the design-variable space (DVS), which consists of 
the particle configurational space as well as the space of lattices via an adaptive fundamental cell, 
the ASC scheme enables one to find 
a point in the DVS in the neighborhood of the starting point that
has a higher packing density than the initial density.
The process is continued until the deepest minimum of 
the objective function (a maximum of packing density) is obtained, which could be either 
a local or global optimum, depending on the particle shapes.

The ASC optimization problem could be solved using 
various techniques, depending on the shapes of the particles. For example, 
for spheres, one can linearize the objective function and the constraints,
and we have found that linear programming techniques can efficiently
produce optimal solutions for such ASC problems \cite{To09b}.
However, for hard polyhedra the non-overlapping conditions given 
by the separation axis theorem (discussed in detail below) 
involve at least quartic inequalities, which makes it inefficient to 
solve even using nonlinear programming methods.

Here, for polyhedral particles, we solve the ASC
optimization problem using a stochastic procedure, i.e., 
Monte Carlo (MC) method with a Metropolis acceptance 
rule for trial moves to search the DVS efficiently. 
However, it is important to distinguish our procedure that 
incorporates deformation and compression/expansion of the 
fundamental cell (i.e., the space of lattices) as design variables from previous 
MC hard-particle packing algorithms \cite{Jo85}. In standard MC 
simulations, arbitrarily selected individual particle is given 
random displacement or rotation. This sequential 
movement method is not able to account for any collective 
motions of the particles, which is crucial to increasing the packing density, 
as pointed out in Sec.~I. In our procedure, the deformation/compression/expansion of the boundary 
at least in part allows for collective particle motions in a direction leading to 
higher packing density. Moreover,
it is the overall compression of the fundamental cell that causes the packing density to increase, 
not the growth of the particles as in most 
MD and MC hard-particle packing algorithms \cite{Do05a,Do05b,Jo85}.

At first glance, one might surmise that an algorithm that employs particle growth
with an adaptive non-shrinking fundamental cell is equivalent to 
our choice of fixing the particle size while allowing the cell
to shrink on average. The former is computationally less efficient
than the latter for polyhedral particles.  Specifically, growth of polyhedral 
particles requires manipulating the coordinates of the vertices of each particle, 
and thus involves at least $dNn_v$ 
numerical operations, where $d$ is the spatial dimension, $N$ is 
the total number of particles in the system and $n_v$ is the number 
of vertices per particle. It is much more computationally 
expensive to use growing/shrinking particles as trial moves 
for the optimization scheme, especially when the number of particles 
is large, compared to our adaptive fundamental 
cell approach, which only requires manipulating $d(d+1)/2$ 
strain components.

In the ASC scheme,
the macroscopic deformation and compression/expansion of the fundamental cell of the lattice 
is completely specified by a strain tensor. Since we only consider small deformations, 
linear strain analysis can be applied here. Starting from an initial configuration, 
a trial configuration can be generated by moving (translating
and rotating) a randomly chosen particle or by a random macroscopic 
deformation and compression/expansion of the fundamental cell. 
If any two particles overlap, the trial configuration is rejected; 
otherwise, if the fundamental cell shrinks in size (which makes the density $\phi$ higher), 
the trial configuration is accepted.
On the other hand,  if the cell expands in size, 
the trial configuration is accepted with a specified probability $p_{acc}$, which 
decreases as $\phi$ increases and approaches zero when the jamming 
limit \cite{To01} (i.e., locally maximal dense packing) is reached. In particular, we find 
$p_{acc}$, with an initial value $p_{acc}\sim 0.35$,  decreasing as a power law with 
exponent equal to -1 works well for 
most systems that  we studied. 
The particle motion is equally likely to be a translation or a rotation. 
The ratio of the number of particle motions to the number of boundary trial moves 
should be greater than unity (especially towards the end of the simulation), 
since compressing a dense packing could
result in many overlaps between the particles. Depending on the initial density, 
the magnitudes of the particle motions and strain components (e.g., the dynamical parameters) need to be 
chosen carefully to avoid the system getting stuck in some shallow local minimum. 
The parameters should also be adjusted accordingly as the packing density increases, 
especially towards the jamming point. 
The number of total Monte-Carlo moves per particle is of the order of $5 \times 10^6$.

In the ensuing subsections, we describe in detail how we implement particle motions as well as 
the deformation and compression/expansion of the fundamental cell, and precise check
for interparticle overlaps using the 
separation axis theorem \cite{Ra94}. We also discuss the cell method and nearest-neighbor lists
that are employed to speed up the simulations.

\subsection{Particle Motions}

The fact that a  polyhedron is the convex hull of its vertices makes
the set of vertices  a useful geometrical representation of a such particle. It is convenient to choose
the origin of the local coordinate system for the vertices to be the centroid of the polyhedron.
Other important geometrical properties of the polyhedron, such as its faces and edges, 
can be represented as certain subsets of the vertices. 

Let the Euclidean position of the centroid of the $j$th particle be ${\bf x}^E_j$ ($j=1,..,N$). 
A translational motion of the particle centroid  can be obtained by generating 
a randomly oriented displacement $\Delta {\bf x}^E_j$ with small magnitude 
($10^{-4} - 10^{-6}$ of the characteristic length of the particle), i.e.,
\begin{equation}
{\bf \bar{x}}^E_j = {\bf x}^E_j+\Delta {\bf x}^E_j.
\label{eq1}
\end{equation}
A rotational motion can be generated by rotating the particle (all of its vertices) 
along a randomly chosen axis (passing through its centroid) by a random small angle 
$\theta$. Let the vector ${\bf v}_i$ originating at the centroid denote the vertex $i$.
We have 
\begin{equation}
{\bf \bar{v}}_i = {\bf R}\cdot{\bf v}^{\perp}_i + {\bf v}^{\parallel}_i,
\label{eq2}
\end{equation}
where ${\bf v}^{\perp}_i$ and ${\bf v}^{\parallel}_i$ are the components of ${\bf v}_i$ 
perpendicular and parallel to the rotation axis, respectively, and ${\bf R}$ is the rotation matrix, i.e., 
\begin{equation}
{\bf R} =\left [{
\begin{array}{c@{\hspace{0.25cm}}c@{\hspace{0.25cm}}c} 
                 \cos\theta & -\sin\theta & 0 \\ 
                 \sin\theta & \cos\theta & 0\\ 
                    0 & 0 & 1 \end{array}}\right]. 
\end{equation}
If this motion does not result in the overlap with another particle, the trial move is accepted; 
otherwise it is rejected.   

\subsection{Adaptive Fundamental Cell}

For a lattice-based periodic packing, the fundamental cell is specified 
by the lattice vectors ${\bf a}_i$ ($i=1,2,3$). Recall that the Euclidean 
coordinates of the particle centroids are ${\bf x}^E_j$ ($j=1,..,N$). 
The relative coordinates of the centroids with 
respect to the lattice vectors are given by
\begin{equation}
{\bf x}^E_j = {\bf \Gamma}\cdot{\bf x}^L_j,
\label{eq3}
\end{equation} 
where ${\bf \Gamma} = [{\bf a}_1, {\bf a}_2, {\bf a}_3]$. 


The adaptive fundamental cell allows for a small strain of the fundamental cell, 
including both volume and shape changes, which is
 represented by a symmetric strain tensor ${\boldsymbol \varepsilon}$, i.e.,
\begin{equation}
\Delta {\bf \Gamma} = {\boldsymbol\varepsilon}\cdot{\bf \Gamma},
\label{eq4}
\end{equation}
 where
\begin{equation}
{\boldsymbol\varepsilon} = \left [{
\begin{array} {c@{\hspace{0.35cm}}c@{\hspace{0.35cm}}c} 
\epsilon_{11} & \epsilon_{12} & \epsilon_{13} \\
\epsilon_{21} & \epsilon_{22} & \epsilon_{23} \\
\epsilon_{31} & \epsilon_{32} & \epsilon_{33}  \end{array} }\right ], 
\end{equation}
and the new fundamental cell (lattice vectors) are given by
\begin{equation}
{\bf \bar{\Gamma}} = {\bf \Gamma} + \Delta {\bf \Gamma}.
\label{eq5}
\end{equation}
Substituting the above equation into Eq.~(\ref{eq1}), we have
\begin{equation}
{\bf \bar{x}}^E_j = {\bf \bar{\Gamma}}\cdot{\bf x}^L_j = {\bf x}^E_j+\Delta {\bf \Gamma}\cdot{\bf x}^L_j.
\label{eq6}
\end{equation}

Thus, the strain of the fundamental cell corresponds to non-trivial 
\textit{collective} motions of the particle centroids (see Fig.~\ref{fig_Cell}). 
In general, the translational motions of the particles contain the 
contributions from a random independent part (given by Eq.~(\ref{eq1})) 
and the collective motion imposed by the adaptive fundamental cell. It is this 
collective motion that enables the algorithm to explore the 
configuration space more efficiently and to produce highly dense packings.

\subsection{Checking Overlaps and the Separation Axis Theorem}

\begin{figure}[bthp]
\begin{center}
$\begin{array}{c}\\
\includegraphics[width=8.5cm,keepaspectratio]{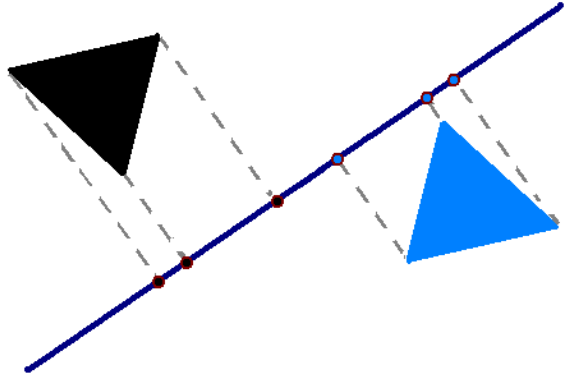} \\
\end{array}$
\end{center}
\caption{(color online). Two nonoverlapping particles and one of their 
separation axes.} 
\label{fig_Axis}
\end{figure}

Hard polyhedron particles, unlike spheres, ellipsoids and superballs, do not possess simple 
overlap potential functions. Thus, the check of overlapping for such particles 
requires other techniques. Two convex objects are separated in space if and only 
if there exist an axis, on which the line segments defined by projections of 
the two objects do not overlap (see Fig.~\ref{fig_Axis}). 
This statement is usually referred to as the {\it separation axis theorem} (SAT) \cite{Ra94}.

For convex polyhedra, the above theorem has a simpler version. 
Since a polyhedron is completely defined by its vertices, 
the line segment on the axis is defined by the most separated two projections of the vertices. 
Moreover, the axis is either perpendicular to one of the faces or 
perpendicular to a pair of edges, one from each polyhedron. 
This reduces the number of axes that need to be  checked from infinity to $[E(E-1)/2+2F]$, 
where $E$ and $F$ is the number of edges and faces of the polyhedron, respectively. 
Here we employ the SAT to check for interparticle overlaps numerically 
up to the highest machine precision.

For polyhedra whose circumscribed and inscribed spheres are well defined, 
two particles are guaranteed to overlap if the centroidal separation is 
smaller than the inscribed diameter and guaranteed not to overlap if the 
centroidal separation is larger than the circumscribed diameter. 
This pre-check dramatically speeds up the simulations starting from configurations with low densities.

\subsection{Cell Method and Near-Neighbor List}

\begin{figure}[bthp]
\begin{center}
$\begin{array}{c}\\
\includegraphics[width=8.5cm,keepaspectratio]{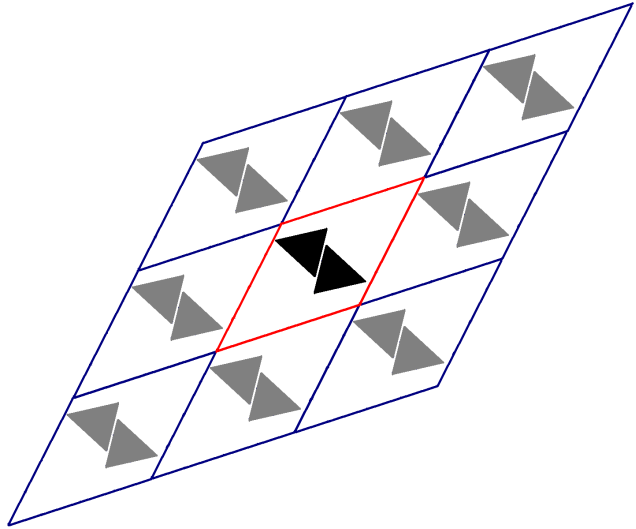} \\
\end{array}$
\end{center}
\caption{(color online). The particles (black) in the central fundamental 
cell (with red boundary) and their images (gray). The distances between 
the particles in the central cell and all the corresponding images need 
to be checked.} 
\label{fig_Periodic}
\end{figure}

The well-developed cell method \cite{Do05a, Do05b} for particle-system simulation is 
used here to speed up the process. However, the simulation box 
(fundamental cell) will not remain a cube during the simulation and 
several conventional techniques developed for cubic box need modifications. 
In particular, to obtain the minimum image distances, 
all surrounding boxes need to be checked explicitly (see Fig.~\ref{fig_Periodic}). 
With the help of the 
relative-coordinate representation of the particle centroids, 
the separation vector of a pair of particles is given by
\begin{equation}
\label{eq7}
{\bf d}_S = {\bf \Gamma}\cdot({\bf x}^L_i-{\bf x}^L_j + {\bf L}), 
\end{equation} 
where ${\bf L}$ is a vector with the values of components to be 1, 0 or -1. 
Note ${\bf L}$ can also be considered as the index of the box that the image particle is in.

The cells are taken to be the same shapes as the simulation box. Partitions
of the particle centroids into the cells are also convenient for relative 
coordinate representations, i.e., the index of the cell 
(represented by a vector) ${\bf C}$ for particle $i$ is given by
\begin{equation}
\label{eq8}
{\bf C} = [N_C {\bf x}^L_i],
\end{equation} 
where $[X]$ gives the smallest integer part of $X$ and $N_C$ is the number of cells. 
Note that the boundary deformations and rotations of particles do not affect 
the cells in which the particles are situated; only translations may cause 
transitions between the different cells.

For dense packings, we make use of a near-neighbor list (NNL) to further improve 
the efficiency of the algorithm \cite{Do05a, Do05b}. In particular, 
when the packing density is high, 
each particle is ``trapped" in a ``cage'' formed by its near neighbors. 
These near-neighbor configurations are practically stationary, i.e.,
the particles undergo very small ``jiggling" motions.  Thus, to check for 
overlapping between the particles, one only needs to consider a particles
nearest neighbors. If a large magnitude translation is made, the NNL needs to be set up again.

\section{Application of the ASC Scheme to the Platonic Solids and Results}

Here we apply the ASC scheme to obtain the densest known packings of the
non-tiling Platonic solids. Due to its lack of central symmetry, the
tetrahedron presents the greatest challenge for the numerical solution 
procedure of the ASC scheme 
because of its tendency to get stuck in local (density) minima, which is a
consequence of the associated ``rugged" energy or, more precisely, ``density" landscape (e.g., the packing 
density as function of the centroidal positions and orientations of all
of the particles in  the fundamental cell). Hence, the
choice of initial configurations becomes crucial in getting dense
tetrahedral packings. By contrast, the central symmetry of the octahedron,
dodecahedron and icosahedron results in density landscapes for the 
numerical procedure that are appreciably less rugged than that for
tetrahedral packings.

\subsection{Tetrahedra}

The determination of the densest packings of regular tetrahedra is  
part of the 18th problem of Hilbert's famous set of problems. It is of interest
to note that the densest (Bravais) lattice packing of tetrahedra
(which requires all of the tetrahedra to have the same
orientations) has the relatively low density $\phi^L_{max} =18/49=0.367\ldots$ and each tetrahedron
touches 14 others \cite{Ho70}. Recently, Conway and Torquato
showed that the densest packings of regular tetrahedra 
 must not be a Bravais lattice packings, and found packings 
with density as large as $\phi \approx 0.72$ \cite{Co06}. One such packing is based upon
the filling of ``imaginary" icosahedra with the densest arrangement of 20 tetrahedra and then 
employing the densest lattice packing of icosahedra.
A slightly higher density was achieved by a perturbation of the so-called ``Welsh" packings with 
density $\phi=17/24 \approx 0.708$ \cite{Co06}.
Using ``tetrahedral" dice, Chaikin {et al.} \cite{Ch07}
experimentally generated jammed disordered packings of such dice
with $\phi \approx 0.75$. Even though the dice are not
perfect tetrahedra (because the vertices and edges are slightly rounded),
these experimental results suggested that the densest
packings of tetrahedra could exceed the  highest densities
reported by Conway and Torquato. Indeed, Chen \cite{Ch08} has recently discovered a periodic
packing of tetrahedra with $\phi=0.7786\ldots$ \cite{conjecture}.
We will call this the ``wagon-wheels" packing because
the basic subunits consist of two orthogonally intersecting
``wagon" wheels. A ``wagon wheel" consists of five contacting
tetrahedra packed around a common edge (see Fig. 1a of Ref. \onlinecite{Co06}).

\begin{table}[ht]
\centering
\caption{Some packing characteristics of certain 
known tetrahedral packings. Here $\phi$ is the packing density, $N$ 
is the number of particles in the fundamental cell, and $\bar{Z}$ is the average contact number
per particle.}
\begin{tabular}{c@{\hspace{0.45cm}}c@{\hspace{0.45cm}}c@{\hspace{0.45cm}}c@{\hspace{0.45cm}}c}
\hline\hline
Name & $\phi$ & Locally Jammed & $N$  & $\bar{Z}$ \\
\hline
Optimal Lattice  \cite{Ho70} & $18/49 \approx 0.367346$    & Yes  & 1  & 14 \\
Uniform  \cite{Co06}       & $2/3 \approx 0.666666$      & Yes  & 2  & 10 \\
Welsh    \cite{Co06}       & $17/24 \approx 0.708333$    & No   & 34 & 25.9 \\
Icosahedral  \cite{Co06}   & 0.716559 & No   & 20 & 20.6 \\
Wagon Wheels \cite{Ch08}            & 0.778615 & Yes   & 18 & 7.1 \\
\hline\hline
\end{tabular}
\label{tab1}
\end{table}

\subsubsection{Initial Conditions}

Because we will use all of the aforementioned tetrahedral packing arrangements
as initial conditions in our numerical solution procedure for the ASC optimization problem, we describe
them now in a bit more detail.
Table \ref{tab1} summarizes some of their packing characteristics, 
including the packing density $\phi$, the number of particles
$N$ in the fundamental cell, the average number 
of contacting particles per particle $\bar{Z}$, and whether the packing
is locally jammed. A packing is locally jammed 
if each particle is locally trapped by its neighbors, i.e., it cannot be translated or 
rotated while fixing the positions and orientations of all the other particles 
\cite{To01,Do07}.

In the optimal lattice packing, each tetrahedron touches 
14 others via edges and vertices. In the uniform packing, 
the tetrahedra are ``locked" in a planer layer, and each particle has 
eight in-plane partial face-to-face contacts and two  edge-to-edge contacts 
(contributed by the two layers above and below), and therefore each particle 
contacts 10 others.
In the icosahedral packing, the tetrahedra have a greater degree
of freedom to move, and indeed it is not locally jammed. 
The 20 tetrahedra inside the imaginary icosahedron are required to meet at its centroid, 
and so each tetrahedron has 20 vertex-to-vertex  contacts. 
Placing the imaginary icosahedra on the sites of its optimal lattice results
in a tetrahedral packing with 12 partial  face-to-face contacts. Thus, there are 8 tetrahedra that have 20
 contacts per particle and 12 tetrahedra with  21 contacts per particle,
and therefore $\bar{Z} = (8\times20+12\times21)/20 = 20.6$.

 The Welsh packing is based on the ``primitive Welsh'' tessellation of
 $\mathbb{R}^3$ into truncated large tetrahedra and small regular tetrahedra \cite{Co06}.
Interestingly, the ``primitive Welsh'' configuration is closely related to the
uniform packing, i.e., it can be constructed by replicating a periodic packing with
two truncated large tetrahedra and two small regular tetrahedra in the fundamental cell.
Each truncated large tetrahedron can be divided into 12 ``Welsh lows'',
and 4 ``Welsh medials," following the notation in Ref.~\onlinecite{Co06}.
The Welsh medials and Welsh lows are tetrahedra with lower symmetry
than regular tetrahedra, which are referred to as the ``Welsh highs."
To construct the Welsh packing, we insert Welsh highs (regular tetrahedra) into
the Welsh low and  medial regions.  We will refer to
the regular tetrahedra that are placed in the Welsh low, medial and high regions as
\textit{lows}, \textit{medials} and \textit{highs}, respectively.
In the Welsh packing, each high makes 4 face-to-face contacts with 4 medials.
Each medial contacts 24 lows and 3 other medials at the centroid of the large
truncated tetrahedron; moreover, each medial makes a face-to-face contact with one 
high. Thus, each medial has 28 contacts.  Each low contacts 23 other lows and 4
medials at the centroid of the large truncated tetrahedron. Thus, in total each
low has 27 contacts. Each truncated large tetrahedron has 12 lows and 4 medials. Thus,
the number of basis particles in the fundamental cell $N= 34$ and 
 $\bar{Z} = (12\times27+4\times28+1\times4)/17 = 440/17 = 25.882352\ldots$.

 In the fundamental cell of the wagon-wheels packing, there are $N = 18$ particles,
 forming two clusters, each of which includes two "wheels" entangled orthogonally  to each other.
 Each cluster has a central connection particle, an ``upper" wheel and a
 ``lower" wheel.  The central particle contacts the other 8 tetrahedra
 through two of its edges. Each of the other 8 tetrahedra has 4 edge
 contacts with particles in its own cluster and 3 partial face-to-face
 contacts with particles from other clusters.
 Thus the average contact number $\bar{Z} = (1\times8+8\times7)/9 = 64/9 = 7.111111\ldots$.
The coordinates of all of the 18 tetrahedra in the fundamental
cell can be found on the authors' website \cite{web}. 

\subsubsection{Dense Packings}

We employ our algorithm to obtain dense packings of tetrahedra
using initial configurations that are based upon the known packings given in Table \ref{tab1} 
as well as certain dilute packings with carefully chosen fundamental cells and 
the number of particles. Note that these dilute packings were not used as initial configurations 
to obtain the results reported in Ref.~\cite{To09a}.  A range of initial densities
is used to yield the largest possible densities in the final states, as summarized in  Table \ref{tab2}.

For initial conditions using the known packings in Table \ref{tab1}, the fundamental 
cell is isotropically expanded with the relative coordinates
of the tetrahedra fixed so that the initial packing is locally unjammed with
a lower density $\phi_{int}$. The highest density of 0.782 obtained from this
subset of initial conditions started from the ``wagon-wheels" packing,
which was the value reported in Ref. \cite{To09a}.

\begin{figure}[bthp]
\begin{center}
$\begin{array}{c@{\hspace{1.0cm}}c}
\includegraphics[width=5.5cm,keepaspectratio,clip=]{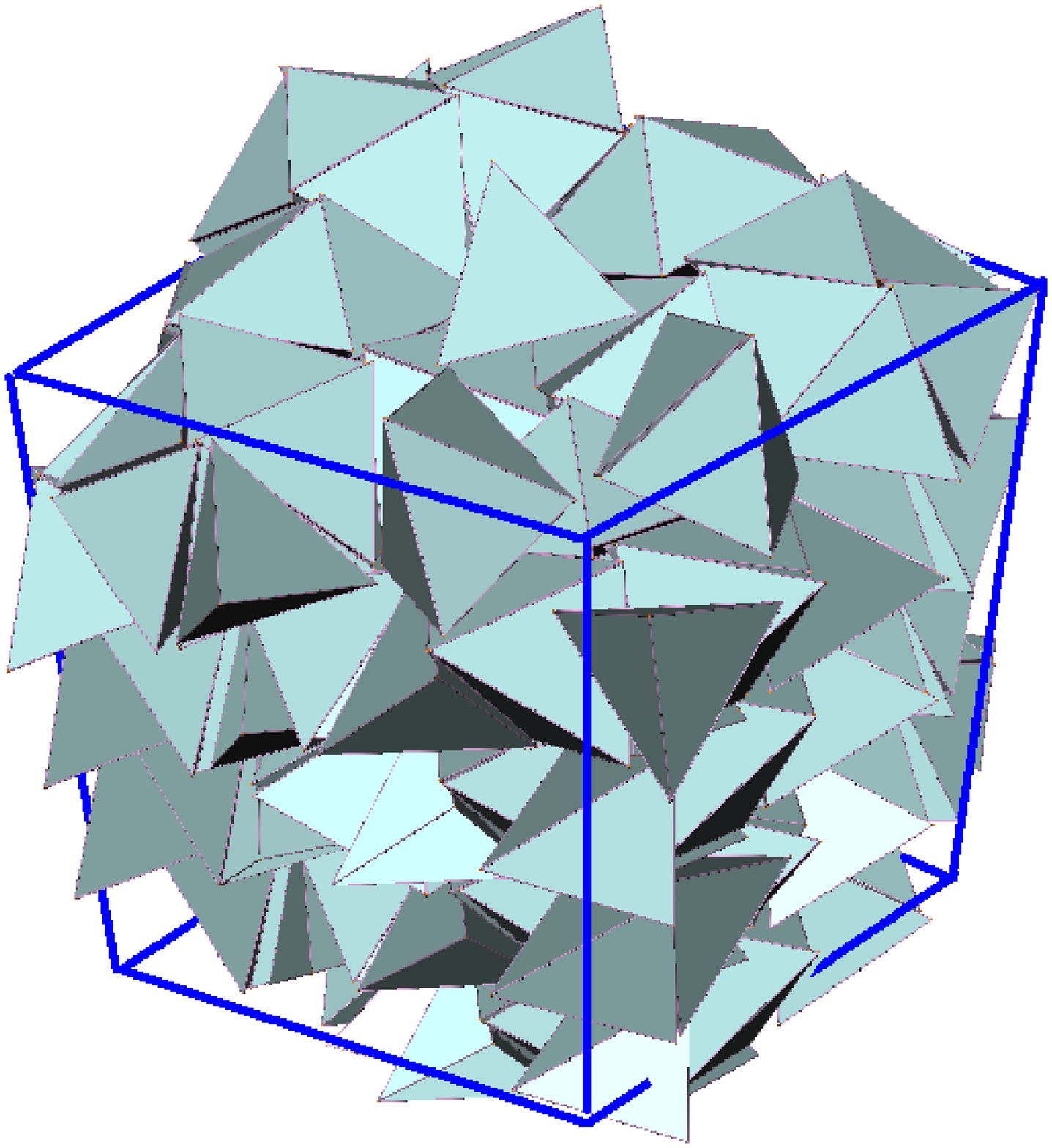} &
\includegraphics[width=5.5cm,keepaspectratio,clip=]{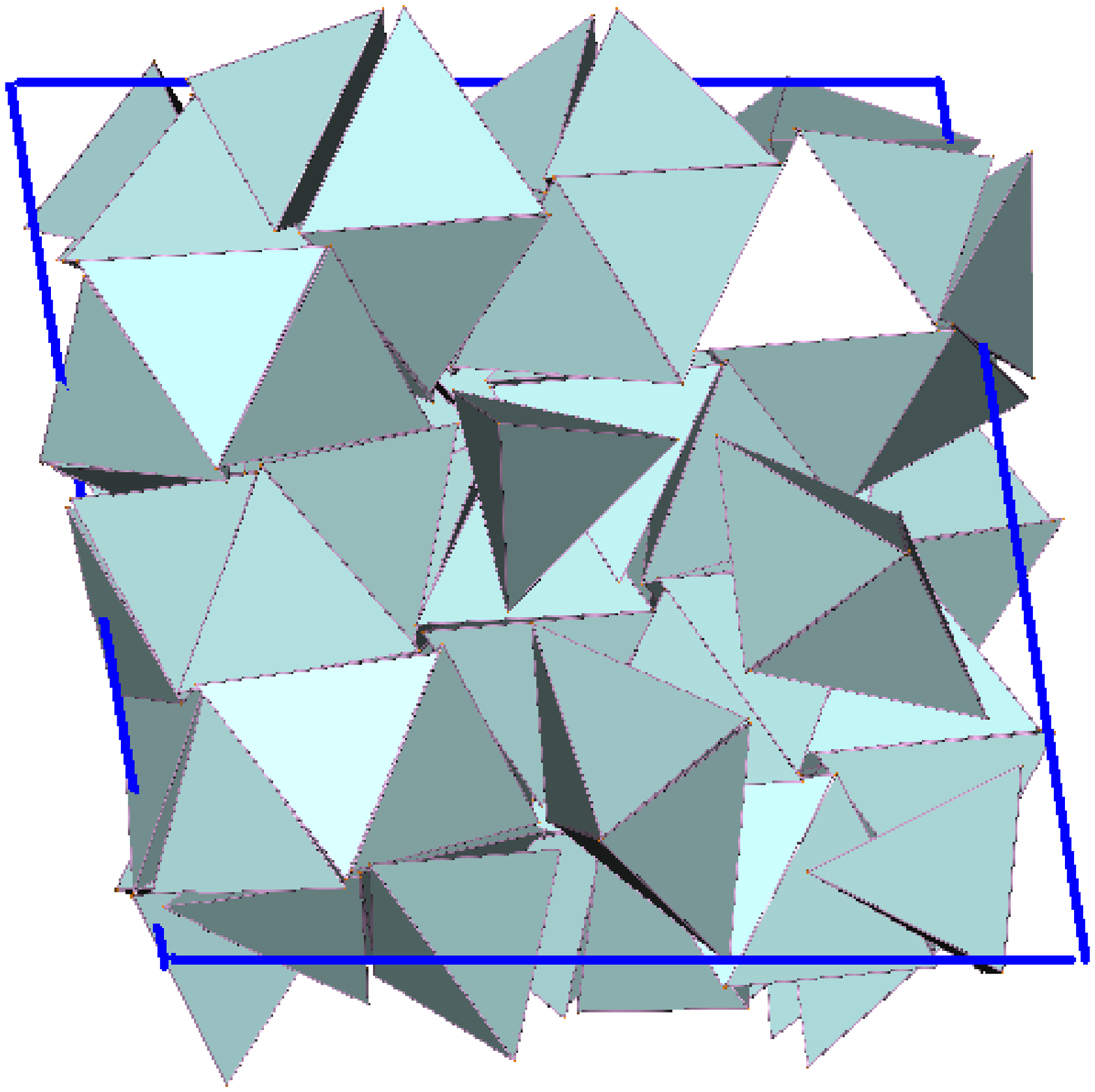} \\
\mbox{(a)} & \mbox{(b)}\\
\end{array}$
\end{center}
\caption{The densest known packing of tetrahedra with 314 particles in the fundamental cell.
We call this the "disordered wagon-wheels" packing. 
(a) Viewed from the side. (b) Viewed from the top.
It is apparent that the packing lacks long-range order
and consists of ``clusters" of distorted wagon wheels.
}
\label{tetra}
\end{figure}

For the aforementioned dilute packings, a variety of fundamental cell shapes spanning from 
that for the simple cubic lattice to that of the hexagonal close packing as well as a
wide range of particle numbers spanning 
from 70 to 350 are explored to generate the dense packings. Specifically, the densest known 
tetrahedral packing is obtained from a dilute initial packing with a rhombical fundamental 
cell similar to that of the hexagonal close packing and 314 particles (see Table \ref{tab2}). 
The particles are originally placed in the fundamental cell randomly and then the system is 
sufficiently equilibrated with the fundamental cell fixed. 
The final packing configuration has a density of about 0.823 
and is shown in Fig. \ref{tetra} from two different viewpoints. The optimized lattice vectors
for the densest packing and other characteristics are given in Appendix B. 
We see that the packing lacks long-range order and is composed of ``'clusters" of  distorted wagon
wheels and individual tetrahedra.  Thus, we call this dense arrangement
the ``disordered wagon wheels" packing. We will comment further
about the significance of achieving this remarkably high density
with a disordered packing in the Discussion and Conclusions (Sec. V).


\begin{table}[ht]
\centering
\caption{Dense tetrahedral packings generated from our algorithm  using initial
configurations based upon the packings given Table \ref{tab1} and the dilute packing as described in the text.
Here $\phi_{int}$ is the initial packing density,
$\phi$ is the final packing density, $N$ is the number
of tetrahedra in the fundamental cell \cite{footnoteTetrah}, 
and  $\bar{Z}$ is the average contact number per particle.}
\begin{tabular}{c@{\hspace{0.45cm}}c@{\hspace{0.45cm}}c@{\hspace{0.45cm}}c@{\hspace{0.45cm}}c@{\hspace{0.45cm}}c}
\hline\hline
Initial Packing & $\phi_{int}$ & $\phi$ & Locally Jammed & $N$  & $\bar{Z}$ \\
\hline
Optimal Lattice & $0.25-0.3$  & $0.695407$   & Yes   & 27  & 6.6 \\
Uniform         & $0.45-0.55$ & $0.665384$    & Yes   & 54  & 9.8 \\
Welsh           & $0.45-0.6$  & $0.752092$    & Yes   & 34  & 7.4 \\
Icosahedral     & $0.45-0.6$  & $0.744532$    & Yes   & 20  & 7.1 \\
Wagon Wheels    & $0.55-0.65$ & $0.782021$    & Yes   & 72  & 7.6 \\
Disordered Wagon Wheels    & $0.005-0.01$ & $0.822637$    & Yes   & 314  & 7.4 \\
\hline\hline
\end{tabular}
\label{tab2}

\end{table}

\subsubsection{Statistics of the Densest Tetrahedral Packing}

\textit{Contact Numbers}: 
The contact numbers of the 
densest tetrahedral packing as well as those obtained using the other five 
initial configurations (Table \ref{tab1}) are 
given in Table \ref{tab2}. 
To determine contacting neighbors, 
it is crucial to find the interparticle 
gaps first. In particular, each of the equilateral triangular faces of the tetrahedra 
is discretized into $M$ equal-sized smaller triangles and 
the distances between the vertices of the small triangles 
of the neighboring tetrahedra are computed, the minimum 
of which is used as the gap value between the two 
corresponding tetrahedra. For our packings, when $M>200$, the 
values of interparticle gaps obtained by this approximation 
scheme become stable and converge to the true gap values. Like any packing 
generated via computer simulations, the 
particles never form perfect contacts and hence contacting neighbors 
must be determined by setting a tolerance $T$ for the interparticle gaps. 
Here we choose $T$ to be equal to the mean value of the gaps 
(of order $10^{-2}$ to $10^{-3}$ of the edge length) such that any gap less than $T$
is associated with a contact. This procedure    
yields the contact numbers given in Table \ref{tab2}. 
In particular, we find that in the disordered wagon wheels packing, 
there are approximately 6.3 face-to-face or partial 
face-to-face contacts and 1.1 edge-to-edge contacts.

\textit{Face-Normal Correlation Function}:
An important statistical descriptor for  packings of 
non-spherical particles is the particle orientation 
correlation function, which measures the extent to which 
a particle's orientation affects the orientation
of another particle at a different position. For highly symmetric
particle shapes, it is reasonable to focus on those  
pair configurations in which the particles are in 
the same orientation, since particle alignment 
is associated with dense configurations and phase transitions that
may occur. However, because tetrahedra lack central symmetry, 
we found that face-to-face contacts  are favored by the dense packings, 
instead of face-to-vertex contacts, which are necessarily associated with aligning 
tetrahedral configurations, such as in the case of the optimal lattice packing.

\begin{figure}[bthp]
\begin{center}
\includegraphics[width=6.0cm,keepaspectratio,clip=]{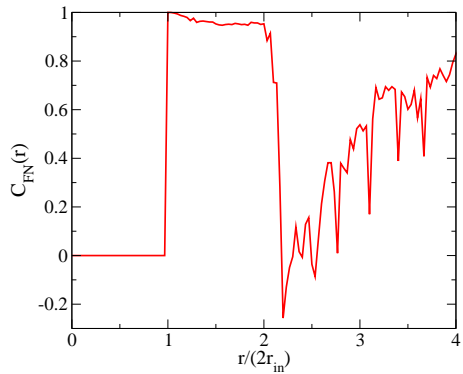}
\end{center}
\caption{The face-normal correlation function $C_{FN}(r)$ of the densest known 
packing of tetrahedra obtained here, where $r_{in}$ is the radius of the insphere of a tetrahedron.} 
\label{fig_CFN}
\end{figure}

Thus, because our interest is in dense tetrahedral
packings, we determine the ``face-normal" correlation function $C_{FN}(r)$, 
which we define as the average of the largest negative value of the 
inner product of two face normals of a pair of tetrahedra separated 
by a distance $r$. This quantity is plotted in Fig.~\ref{fig_CFN}. 
The strong positive peak values, ranging from 
$r = 2r_{in} = d_0/\sqrt{6}$ 
(where $r_{in}$ is the radius of the insphere of the tetrahedron and 
$d_0$ is the edge length of the particle) to $r$ slightly greater than $2r_{in}$,
indicate a large number of face-to-face contacts between neighboring 
particles, implying strong short-range orientational correlations. 
The fact that $C_{FN}(r)$ plateaus to its maximum or near maximum
value, for the distance interval $r/(2r_{in}) \in (1, 2.2)$,
indicates that there are partial face-to-face contacts in which
particles slide relative to one another such that the distance between 
the centroids can increase while the face-normal inner product remains 
the same. The ``valleys" of $C_{FN}(r)$ with relatively small magnitudes 
are manifestations of weak particle alignments (alignment
in the same direction), which are local configurations that
are necessary ``costs'' to achieve the largest face-to-face contact numbers on average.

\begin{figure}[bthp]
\begin{center}
\includegraphics[width=6.0cm,keepaspectratio,clip=]{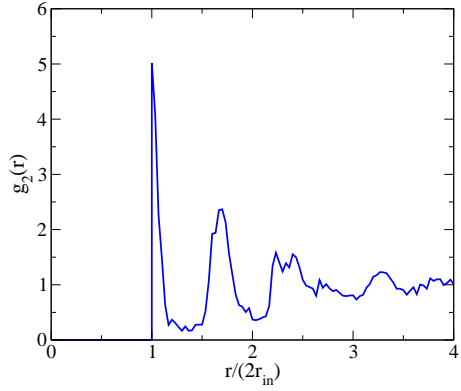}
\end{center}
\caption{The centroidal radial distribution function $g_2(r)$ of the densest known 
packing of tetrahedra, where $r_{in}$ is the radius of the insphere of a tetrahedron.} 
\label{fig_g2}
\end{figure}

\textit{Centroidal Radial Distribution Function}:
Another statistical descriptor of the structure is the centroidal 
radial distribution function $g_2(r)$. In particular, $g_2(r) r^2dr$ is 
proportional to the conditional probability that a particle centroid 
is found in a spherical shell with thickness $dr$ at a radial distance $r$ 
from another particle centroid at the origin.
It is well established that when there is no long-range 
order in the system, $g_2$ decays to unity very fast. 
Figure~\ref{fig_g2} shows the centroidal radial distribution of the 
densest known tetrahedral packing. We see that $g_2(r)$ decays 
to unity after a few oscillations, indicating that the packing lacks
long-range order.

\subsection{Octahedra}
To obtain dense packings of octahedra,  we use a wide range of initial configurations.
This includes unsaturated packings in which the particles are randomly oriented and positioned 
with densities spanning the range from $0.2$ to $0.35$ as well as 
unsaturated simple cubic (SC) and face-centered cubic (FCC) lattice 
packings with densities from 0.1 to 0.125, and the optimal lattice packings 
with densities from 0.3 to 0.55 are used. We found that, using small 
compression rates for the random initial configurations and moderate 
compression rates for the various lattice configurations along with a sufficient 
number of particle displacements and rotations, resulted in final configurations 
with densities larger than $0.93$ that are very close in structure and density 
to the optimal lattice packing. More specifically, 
starting from the unsaturated SC lattice (with density $0.1$) and 
optimal lattice  (with density $0.55$) packings, final packings with 
densities very slightly larger than the value $0.947003\ldots$ can be obtained,
which are extremely close in structure and density to the optimal lattice packing 
($\phi^L_{max} = 0.947368\ldots$) \cite{Mi04}. A previous study
involving an event-driven molecular dynamics growth algorithm
for the case of octahedral-like superballs led to the
same the densest lattice packing, which lends further credence
to the fact that this lattice is indeed optimal among all packings \cite{Ji09}.
 Each octahedron in the 
optimal lattice packing, depicted in Fig. \ref{octa},  contacts 14 others and its lattice vectors
are given in Appendix A.

\begin{figure}[bthp]
\begin{center}
\includegraphics[width=5.5cm,keepaspectratio]{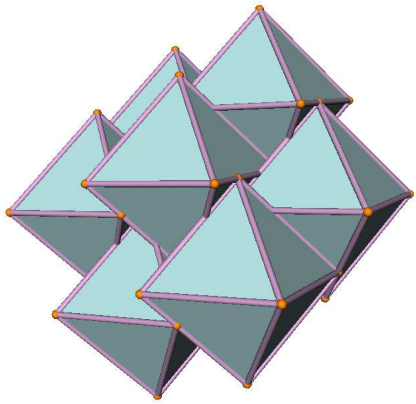}
\end{center}
\caption{A portion of the optimal lattice packing of octahedra.}
\label{octa}
\end{figure}

\subsection{Icosahedra}

As in the case of octahedra, we use a variety of initial conditions
to get dense packings of icosahedra.
This includes unsaturated packings 
in which the icosahedra are randomly
oriented and positioned with densities spanning the range from 0.2 to 0.3 
as well as various unsaturated lattice packings, such as the BCC, 
FCC and the densest lattice packing, with densities spanning the
range from 0.3 to 0.65. We found that by starting from a low-density
configuration, with a small enough compression rate and sufficient
number of particle moves, one can always obtain a final configuration
that is very close to the optimal lattice packing with a density
$\phi \approx 0.83$. An initial configuration
of an unsaturated optimal lattice packing configuration with a density 0.65 gives a
final packing with density  $0.836315 \ldots$, which is extremely close
in structure and density to the optimal lattice packing ($\phi^L_{max} = 0.836357\ldots$) \cite{Be00}. 
Each icosahedron of the optimal lattice packing, depicted in Fig. \ref{icosa},  
contacts 12 others and its lattice vectors
are given in Appendix A.

\begin{figure}[bthp]
\begin{center}
\includegraphics[width=5.5cm,keepaspectratio]{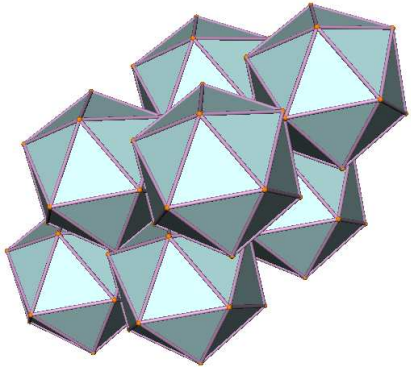}
\end{center}
\caption{A portion of the optimal lattice packing of icosahedra.}
\label{icosa}
\end{figure}

\subsection{Dodecahedra}

Unsaturated random packings with densities ranging from $0.15$ to $0.3$ 
as well as unsaturated simple cubic and the optimal lattice 
packings with densities spanning the range from 0.3 to 0.6 are employed 
as initial configurations to generate dense packings of dodecahedra. 
We found that it is algorithmically more difficult to avoid local density maxima 
for dodecahedra than octahedron and icosahedron packings discussed above.
For example, starting from random configurations, 
even with sufficiently small compression rate and a large number 
of particle moves, we can only achieve apparently jammed final 
packings with densities in the range from 0.83 to 0.85. When we use an
unsaturated optimal lattice packing with density $0.72$ as an initial condition, we can 
generate a final packing with $\phi = 0.904002\ldots$, which  is 
relatively close in structure and density to the optimal lattice packing
($\phi^L_{max} = 0.904508\ldots$) \cite{Be00}. Each
dodecahedron of the optimal lattice packing, depicted in Fig. \ref{dodeca},  contacts 12 others and its
lattice vectors are given in Appendix A.

\begin{figure}[bthp]
\begin{center}
\includegraphics[width=5.5cm,keepaspectratio]{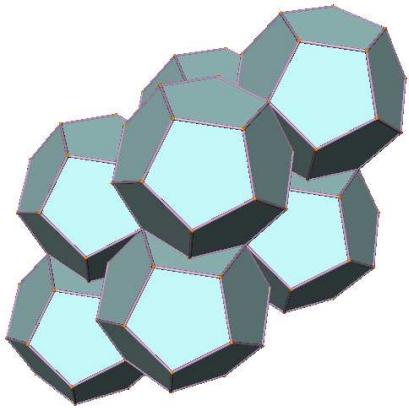}
\end{center}
\caption{A portion of the optimal lattice packing of dodecahedra.}
\label{dodeca}
\end{figure}

We note that the tendency for dodecahedral packings to get stuck in 
local-density maxima in our algorithm is due to the fact that in some
sense the dodecahedron is a  shape that is intermediate between the octahedron
and icosahedron.  Specifically, for octahedral packings, the octahedral symmetry
of the particles facilitates
the formation of nematic phases at relatively high densities, 
which can then be easily compressed into 
a dense crystalline phase with adaptive fundamental cells. 
An icosahedron is highly isotropic and possesses a large 
number of faces. Thus, at low densities, the icosahderal packing behaves like 
a hard-sphere fluid, and  it is only near the jamming point that the 
asphericity of the icosahedron begins to play an important role.
Again, by allowing the fundamental cell to adapt in the case of icosahedra, the optimal 
lattice that best accommodates the packing can be easily identified.
However, for our algorithm, the dodecahedral packing
behaves differently from either the octahedral or icosahedral
packings.  In particular, while dodecahedra are
more isotropic than octahedra, which strongly suppresses
the formation of nematic phases, they have fewer but
larger faces than icosahedra, which favors the formation
of face-to-face contacts that ends up jamming the dodecahedral packing
at lower densities than either the octahedral or icosahedral
packings.

\section{Upper Bound on the Maximal Density of Packings
of Nonspherical Particles}
\label{upper}

Here we derive a simple upper bound on the maximal density
$\phi_{max}$ of a packing of congruent nonspherical particles
of volume $v_p$ in any Euclidean space dimension $d$. We will see
that this bound will aid in our analysis of the optimality
of not only the Platonic and Archimedean solids but also superballs. 
Let $\phi_{max}^S$ be the maximal density of a $d$-dimensional packing
of congruent spheres and let $v_s$ represent
the volume of the largest sphere than can be inscribed
in the nonspherical particle.
\smallskip

\noindent{Lemma:} {\sl The maximal density of a packing of congruent
nonspherical particles
is bounded from above according to the following bound \cite{connelly}:}
\begin{equation}
\phi_{max}\le \mbox{min}\left[\frac{v_p}{v_s}\; \phi_{max}^S,1\right],
\label{lemma}
\end{equation}
where $\mbox{min}[x,y]$ denotes the minimum of $x$ and $y$. 

The proof is straightforward. The maximal packing density $\phi_{max}$
can be expressed in terms of the maximal number density $\rho_{max}$
via the relation
\begin{equation}
\phi_{max}= \rho_{max} v_p.
\end{equation}
If we inscribe within each nonspherical
particle of the packing the largest possible sphere, it is clear that
\begin{equation}
\rho_{max} v_s \le \phi_{max}^S,
\end{equation}
and therefore combination of the last two equations yields
the upper bound of the Lemma.

\noindent{\it Remark:} The upper bound (\ref{lemma}) will yield a
reasonably tight bound for packings of nonspherical particles provided that
the {\it asphericity} $\gamma$  of the particle is not large. Here we define
the asphericity as the following ratio:
\begin{equation}
\gamma=\frac{r_{out}}{r_{in}}
\end{equation}
where $r_{out}$ and $r_{in}$ are the circumradius and inradius
of the circumsphere and insphere of the nonspherical particle.
The circumsphere is the smallest sphere containing the particle.
The insphere is the largest sphere than can be inscribed
in the particle. For a sphere, clearly the asphericity $\gamma=1$.
Since upper bound (\ref{lemma}) cannot be sharp (i.e., exact) for a nonspherical
particle, any packing construction for a nonspherical
particle whose density is close to the upper bound
(\ref{lemma}) is nearly optimal, if not optimal.

In the three-dimensional case, the upper bound (\ref{lemma})
becomes
\begin{equation}
\phi_{max}\le \phi_{max}^U = \mbox{min}\left[\frac{v_p}{v_s}\; \frac{\pi}{\sqrt{18}},1\right].
\label{bound}
\end{equation}
We now apply this upper bound to packings of
the Platonic and Archimedean solids and compare the bounds to the densities of the corresponding densest lattice packings. Moreover, we apply the upper bounds
to superball and ellipsoid packings.

\subsection{Platonic Solids}

Table \ref{plat} compares the density of the densest
lattice packings of the Platonic solids
to the corresponding upper bounds on the maximal density for such
packings. The large asphericity and lack of central symmetry of the
tetrahedron is consistent with the large gap between the upper
bound density and densest lattice packing density, and the
fact that there are non-lattice packings with density appreciably
greater than $\phi_{max}^L$. On the other hand, the central
symmetry of the octahedron, dodecahedron and icosahedron and
their associated relatively small asphericities explain
the corresponding small differences between $\phi_{max}^L$ and $\phi_{max}^U$
and is consistent with our simulation findings that
indicate that their optimal arrangements are their respective densest
lattice packings.

\begin{table}[ht]
\centering
\caption{Comparison of the densities of the densest lattice packings for
the Platonic solids \cite{Mi04,Ho70,Be00} to the
corresponding upper-bound densities as obtained from (\ref{bound}).
Here $v_p$ is volume of the polyhedron with \textit{unit} edge length,
$r_{in}$ and $r_{out}$ are the radii of the insphere and circumsphere of the polyhedron with \textit{unit}
edge length, respectively, $\gamma = r_{out}/r_{in}$ is the asphericity, $\phi_{max}^L$ is the density of
the optimal lattice packing and $\phi_{max}^U$ is the upper bound
(\ref{bound}). The numerical values are reported up to the sixth decimal place. The naming code 
used here is the same one used in Fig.~1.}
\begin{tabular}{c@{\hspace{0.25cm}}c@{\hspace{0.25cm}}c@{\hspace{0.25cm}}c@{\hspace{0.25cm}}c@{\hspace{0.25cm}}c@{\hspace{0.25cm}}c}
\hline\hline
Name & $v_p$ & $r_{in}$ & $r_{out}$ &
$\gamma$ & $\phi_{max}^L$ & $\phi_{max}^U$ \\
\hline
Tetrahedron (P1) & $\frac{\sqrt{2}}{12}$ & $\frac{\sqrt{6}}{12}$ & $\frac{\sqrt{6}}{4}$ &
$3$ & $0.367346$ & $1$\\
Icosahedron (P2) & $\frac{5(3+\sqrt{5})}{12}$ & $\frac{3\sqrt{3}+\sqrt{15}}{12}$ & $\frac{\sqrt{10+2\sqrt{5}}}{4}$ &
$1.258410$ & $0.836357$ & $0.893417$\\
Dodecahedron (P3) & $\frac{15+7\sqrt{5}}{4}$ & $\frac{\sqrt{250+110\sqrt{5}}}{20}$ & $\frac{\sqrt{3}+\sqrt{15}}{4}$ & 
$1.258410$ & $0.904508$ & $0.981162$\\
Octahedron (P4) & $\frac{\sqrt{2}}{3}$ & $\frac{\sqrt{6}}{6}$ & $\frac{\sqrt{2}}{2}$ &
$1.732050$ & $0.947368$ & $1$ \\
Cube (P5) & $1$ & $\frac{1}{2}$ & $\frac{\sqrt{3}}{2}$ &
$1.732050$ & $1$ & $1$ \\
\hline\hline
\end{tabular}
\label{plat}
\end{table}

\subsection{Archimedean Solids}

We also compute the upper bound (\ref{bound}) for each of the 13
Archimedean solids and compare them to the densities
of the corresponding densest lattice packings \cite{Ho70,Be00}.
(In Appendix A we provide the lattice vectors for the optimal
lattice packings of the Archimedean solids).
Table \ref{arch} summarizes the upper bounds on the maximal density for
such packings. Not surprisingly, the truncated tetrahedron
(the only Archimedean solid that is not centrally symmetric)
has a large asphericity, implying that there are
denser non-lattice packings, as we explicitly identify
in Section \ref{discussion}. The central symmetry of the majority of the Platonic and Archimedean solids
and their associated relatively small asphericities
explain the corresponding small differences between $\phi_{max}^L$
and $\phi_{max}^U$ and is consistent with our simulation findings
that strongly indicate that their optimal arrangements are their respective
densest lattice packings.

\begin{table}[ht]
\centering
\caption{Comparison of the densities of the densest lattice packings for
the Archimedean solids to the
corresponding upper-bound densities as obtained from (\ref{bound}).
Except for the cubeoctahedron \cite{Ho70}, the densities of the densest
lattice packings were obtained by Betke and Henk \cite{Be00}.
Here $v_p$ is volume of the polyhedron with \textit{unit} edge length,
$r_{in}$ and $r_{out}$ are the radii of the insphere and circumsphere 
of the polyhedron with \textit{unit}
edge length, respectively, $\gamma = r_{out}/r_{in}$ is the asphericity, 
$\phi_{max}^L$ is the density of the optimal
lattice packing, $\phi_{max}^U$ is the upper bound (\ref{bound}) and
$t = [1+(19-3\sqrt{33})^{\frac{1}{3}}+(19+3\sqrt{33})^{\frac{1}{3}}]/3
\approx 1.83929\ldots$ is the Tribonacci constant. The numerical 
values are reported up to the sixth decimal place. The naming code used here 
is the same one used in Fig.~3.}
\begin{tabular}{c@{\hspace{0.15cm}}c@{\hspace{0.15cm}}c@{\hspace{0.15cm}}c@{\hspace{0.15cm}}c@{\hspace{0.25cm}}c@{\hspace{0.25cm}}c}
\hline\hline
Name & $v_p$ & $r_{in}$ & $r_{out}$ &
$\gamma$ & $\phi_{max}^L$ & $\phi_{max}^U$ \\
\hline
Truncated Tetrahedron (A1) & $\frac{23\sqrt{2}}{12}$ & $\frac{\sqrt{6}}{4}$ & $\frac{\sqrt{22}}{4}$ &
$1.914854$ & $0.680921$ & $1$\\
Truncated Icosahedron (A2) & $\frac{125+43\sqrt{5}}{4}$ & $\frac{\sqrt{21+9\sqrt{5}}}{2\sqrt{2}}$ & $\frac{\sqrt{58+18\sqrt{5}}}{4}$ &
$1.092945$ & $0.784987$ & $0.838563$ \\
Snub Cube (A3)& $\frac{3\sqrt{t-1}+4\sqrt{t+1}}{3\sqrt{2-t}}$ & $\sqrt{\frac{t-1}{4(2-t)}}$ & $\sqrt{\frac{3-t}{4(2-t)}}$ &
$1.175999$ & $0.787699$ & $0.934921$\\
Snub Dodecahedron (A4)& $37.616654$ & $1.980915$ & $2.155837$ &
$1.088303$ & $0.788640$ & $0.855474$ \\
Rhombicicosidodecahedron (A5)& $95+50\sqrt{5}$ & $3.523154$ & $\frac{\sqrt{31+12\sqrt{5}}}{2}$ &
$1.079258$ & $0.804708$ & $0.835964$\\
Truncated Icosidodecahedron (A6)& $\frac{60+29\sqrt{5}}{3}$ & $2.016403$ & $\frac{\sqrt{11+4\sqrt{5}}}{2}$ &
$1.107392$ & $0.827213$ & $0.897316$\\
Truncated Cuboctahedron (A7)& $\frac{12+10\sqrt{2}}{3}$ & $\frac{1+\sqrt{2}}{2}$ & $\frac{\sqrt{5+2\sqrt{2}}}{2}$ &
$1.158941$ & $0.849373$ & $0.875805$\\
Icosidodecahedron (A8)& $\frac{45+17\sqrt{5}}{6}$ & $\sqrt{\frac{5+2\sqrt{5}}{5}}$ & $\frac{1+\sqrt{5}}{2}$ &
$1.175570$ & $0.864720$ & $0.938002$\\
Rhombicuboctahedron (A9)& $22+14\sqrt{2}$ & $\frac{1+2\sqrt{2}}{2}$ & $\frac{\sqrt{13+6\sqrt{2}}}{2}$ &
$1.210737$ & $0.875805$ & $1$ \\
Truncated Dodecahedron (A10)& $\frac{5(99+47\sqrt{5})}{12}$ & $\frac{\sqrt{25+11\sqrt{5}}}{2\sqrt{2}}$ & $\frac{\sqrt{74+30\sqrt{5}}}{4}$ &
$1.192598$ & $0.897787$ & $0.973871$\\
Cuboctahedron (A11)& $\frac{5\sqrt{2}}{3}$ & $\frac{\sqrt{2}}{2}$ & $1$ &
$1.414213$ & $0.918367$ & $1$\\
Truncated Cube (A12)& $\frac{21+14\sqrt{2}}{3}$ & $\frac{1+\sqrt{2}}{2}$ & $\frac{\sqrt{7+4\sqrt{2}}}{2}$ &
$1.473625$ & $0.973747$ & $1$\\
Truncated Octahedron (A13)& $8\sqrt{2}$ & $\frac{\sqrt{6}}{2}$ & $\frac{\sqrt{10}}{2}$ &
$1.290994$ & $1$ & $1$ \\
\hline\hline
\end{tabular}
\label{arch}
\end{table}

\subsection{Superballs}

The upper bound (\ref{bound}) in the case of superballs \cite{Ji09}
can be expressed analytically for all values of the deformation parameter
$p$. A three-dimensional \textit{superball} is a centrally symmetric body in
$\mathbb{R}^3$
occupying the region
\begin{equation}
|x_1|^{2p}+|x_2|^{2p}+|x_3|^{2p} \le 1,
\end{equation}
where $x_i$ $(i=1,2,3)$ are Cartesian coordinates
and $p \ge 0$ is the \textit{deformation parameter}, which
indicates to what extent the particle shape has deformed from that
of a sphere ($p=1$). A superball can possess two types of shape
anisotropy: cubic-like shapes for $1 \le p \le \infty$, with
$p=\infty$ corresponding to the perfect cube,
and octahedral-like shapes for $0 \le p \le 1$,
with $p=1/2$ corresponding to the perfect regular octahedron.
In Ref. \onlinecite{Ji09}, event-driven molecular dynamics growth algorithms
as well as theoretical arguments led to the conjecture
that the densest packings of superballs for all convex
shapes ($1/2 \le p \le \infty$) are certain lattice packings,
depending on the value of the deformation parameter $p$.

For convex superballs in the octahedral regime
($0.5 < p < 1$), the upper bound (\ref{bound}) is explicitly given by
\begin{equation}
\displaystyle{\phi_{max}^U = \frac{\sqrt{6}}{108 p^2}\times
3^{\frac{1}{2p}} B\left
({\frac{1}{2p},\frac{2p+1}{2p}}\right)B\left
({\frac{1}{2p},\frac{p+1}{p}}\right)},
\end{equation}
similarly, for superballs in the cubic regime ($p>1$),
the upper bound (\ref{bound}) is given by
\begin{equation}
\displaystyle{\phi_{max}^U = \frac{\sqrt{2}}{4p^2} B\left
({\frac{1}{2p},\frac{2p+1}{2p}}\right)B\left
({\frac{1}{2p},\frac{p+1}{p}}\right)},
\end{equation}
where $B(x,y) = \Gamma(x)\Gamma(y)/\Gamma(x+y)$ and $\Gamma(x)$ is the
Euler Gamma function. Here and in the subsequent paragraphs
concerning ellipsoids, we only state the nontrivial part of the 
upper bound (\ref{bound}).
For $p$ near the sphere point ($p=1$), the densest lattice packings
\cite{Ji09} have densities
that lie relatively close to the corresponding upper-bound values. For example,
for $p = 0.99, 0.98$ and 0.97, $\phi_{max}^U = 0.745327\ldots,
0.750274\ldots$,
and $ 0.755325\ldots$, respectively, which is to be compared to
$\phi_{max}^L= 0.740835\ldots, 0.741318\ldots$, and $0.741940\ldots$,
respectively.
For $p = 1.01, 1.02$ and 1.03, $\phi_{max}^U = 0.747834\ldots,
0.755084\ldots$,
and $0.762233\ldots$, respectively, which is to be compared to
$\phi_{max}^L=0.741720\ldots, 0.742966\ldots$, and $0.744218\ldots$,
respectively.

The fact that the constructed densest lattice packings of superballs have density 
$\phi_{max}^L$ relatively close to the upper bound value $\phi_{max}^U$ 
strengthens the arguments made in Ref.~\onlinecite{Ji09} that suggested that the 
optimal packings are in fact given by these arrangements. If the optimal 
packings around $p=1$ are indeed the lattice packings, it would be 
surprising that for other values of $p$ a continuous deformation of the 
superballs would result in a transition from lattice packings
to denser non-lattice packings.

\subsection{Ellipsoids}

In the case of ellipsoids in which the
the ratio of the three semiaxes are given by $1:\alpha:\beta$ ($\alpha,
\beta \ge 1$),
the upper bound (\ref{bound}) becomes
\begin{equation}
\displaystyle{\phi_{max}^U = \alpha\beta \frac{\pi}{\sqrt{18}}}.
\end{equation}
For prolate spheroids $\beta = 1$, and the bound becomes
\begin{equation}
\displaystyle{\phi_{max}^U = \alpha \frac{\pi}{\sqrt{18}}}.
\end{equation}
For oblate spheroids with $\alpha = \beta > 1$, the bound is given by
\begin{equation}
\displaystyle{\phi_{max}^U = \alpha^2\frac{\pi}{\sqrt{18}}}.
\end{equation}
The upper bounds for ellipsoids generally do not work as well as those
for the centrally symmetric Platonic and Archimedean
solids, and superballs.
This is due to the fact that the three principle directions (axes) are not
equivalent for an ellipsoid, i.e., it generally possesses different semiaxes. 
Note that the three equivalent principle (orthogonal) axes  (directions)
of a centrally symmetric particle are those directions that are 
two-fold rotational symmetry axes  such that the
distances from the particle centroids to the particle surfaces are equal.
In addition, the asphericity of an ellipsoid increases linearly as the 
largest aspect ratio $\alpha$ increases without limit.  By
contrast, the three principle directions are
equivalent for the centrally-symmetric Platonic and Archimedean solids as
well as for superballs, for which the asphericity is always bounded and close 
to unity (see Tables \ref{plat} and \ref{arch}).
Thus, we see that an asphericity value close to unity 
is a necessary condition to have a  centrally symmetric particle
in which the three principle directions are equivalent. 


\section{Discussion and Conclusions}
\label{discussion}

We have formulated the problem of generating dense packings of 
nonoverlapping polyhedra  within an adaptive fundamental cell  
subject to periodic boundary conditions as  an optimization problem,
which we call the Adaptive Shrinking Cell (ASC) scheme.
The  procedure allows both a sequential search of the configurational space of 
the particles and the space of lattices via an adaptive fundamental cell that shrinks
on average to obtain dense packings. We have applied the ASC to generate the
densest known packings of the Platonic solids.

For tetrahedra, we find a packing with density $\phi\approx 0.823$,
which  is a periodic (non-Bravais lattice) packing  with a complex
basis.
Unlike the other Platonic solids, finding dense packings of tetrahedra
with our algorithm requires having good initial configurations. The densest
packing was found using 314 particles in a rhombical fundamental cell 
that is similar to that of the hexagonal close packing.
As we stressed in our earlier work \cite{To09a} and
continue to confirm in this paper, it is possible
that denser packings of tetrahedra will involve increasingly larger 
numbers of particles in the fundamental cell. In fact, the higher density 
found here is realized by a larger periodic packing (314 particles per cell) than 
the one reported in Ref.~\cite{To09a} (72 particles per cell) with density $0.782$. 
It is apparent the obtained densest known tetrahedral packing is 
disordered in the sense that it possesses no long-range order,
at least on the scale of the simulation box.
 This packing can be considered to be a disordered ``mixture'' of distorted 
wagon wheels and individual tetrahedra and it is distinct from the one reported in 
Ref.~\cite{To09a} in both the arrangement of the wagon wheels and 
the extent of the distortion of the wagon wheels. Although we cannot  
rule out the possibility of the existence of denser packings involving a more 
ordered arrangement of the wagon wheels and individual tetrahedra, it 
is reasonable to expect that wagon wheels will be key building blocks
in denser packings and should at least be slightly distorted to 
fill the interparticle gaps in wagon-wheel clusters more efficiently, which necessarily 
introduces a certain 
degree of disorder to the packing. If denser packings of tetrahedra must
involve larger number of particles without long-range order, than it
raises the amazing prospect that the densest packings of tetrahedra
might be truly disordered, due to the geometrical frustration
associated with the lack of central symmetry of a
tetrahedron and that tetrahedra cannot tile space. This would
be the first example of a maximally dense packing of
congruent convex three-dimensional particles without long-range order.
However, we cannot offer definitive conclusions about
this possibility at this stage. It is clear that in future work in the search for denser packings,
increasingly larger number of tetrahedra must be considered,
which can only be studied using greater computational resources.

Our simulation results and rigorous upper bounds strongly suggest that the optimal lattice 
packings of the centrally symmetric
Platonic solids (octahedra, dodecahedra and icosahedra) are indeed the 
densest packings of these particles, especially since these
arise from a variety of initial ``dilute'' multi-particle 
configurations within the fundamental cell \cite{footnoteBarlow}. It is noteworthy that
a different simulation procedure (an event-driven molecular
dynamics growth algorithm with an adaptive fundamental cell) has recently been 
used to demonstrate that the densest packings of octahedral-like superballs  
(which contains the perfect octahedron) are likely to be the optimal lattice packings \cite{Ji09}. 
Moreover, the fact that the optimal lattice 
packing densities of certain centrally symmetric nonspherical particles, 
such as the Archimedean solids with central symmetry and convex superballs, 
are relatively close to their upper bounds as well as other theoretical arguments
given below suggest that the densest packings of these particles 
may also be given by their optimal lattice packings.

It is crucial to stress that the nonspherical particles in this 
family do not deviate appreciably from a sphere, i.e.,  
their corresponding asphericity $\gamma$
is always bounded and relatively close to unity, and, moreover, their three principle axes
(directions) are equivalent. This is in contrast to the
ellipsoid, which, although centrally symmetric, generally possesses 
three principle axes (directions) that are inequivalent and  its asphericity 
can increase without limit as the largest aspect ratio grows. These characteristics
suggest that the densest ellipsoid arrangements are non-lattice
packings, which indeed has been verified \cite{Do04}.


Our simulation results and the ensuing theoretical arguments 
lead to the following conjecture:
\smallskip

\noindent{Conjecture 1:} {\sl The densest packings of the centrally symmetric 
Platonic solids are given by their corresponding optimal lattice packings.}

\begin{figure}[bthp]
\begin{center} 
$\begin{array}{c@{\hspace{0.25cm}}c@{\hspace{0.25cm}}c@{\hspace{0.25cm}}c}\\
\includegraphics[height=3.0cm,keepaspectratio]{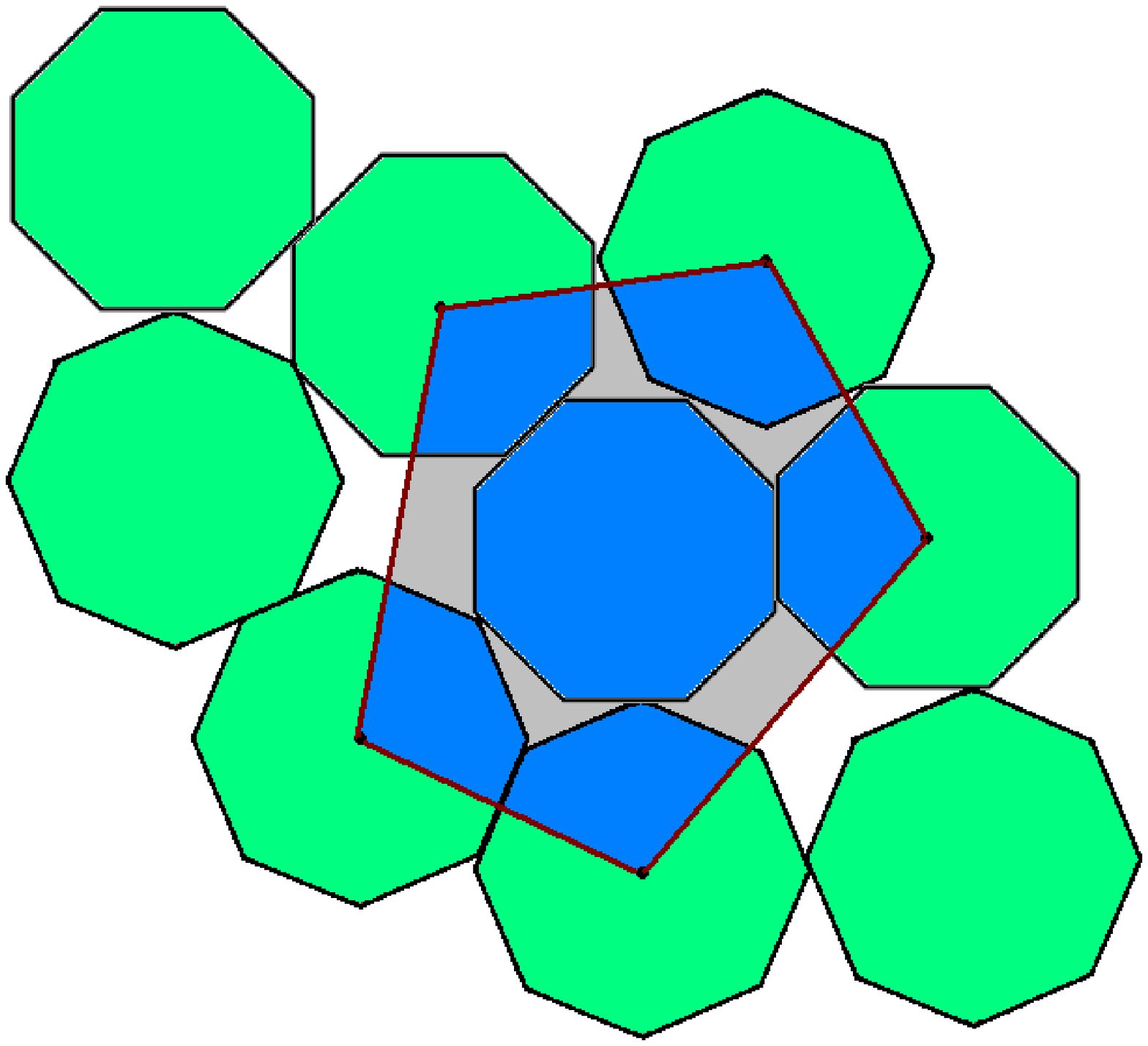} &
\includegraphics[height=3.0cm,keepaspectratio]{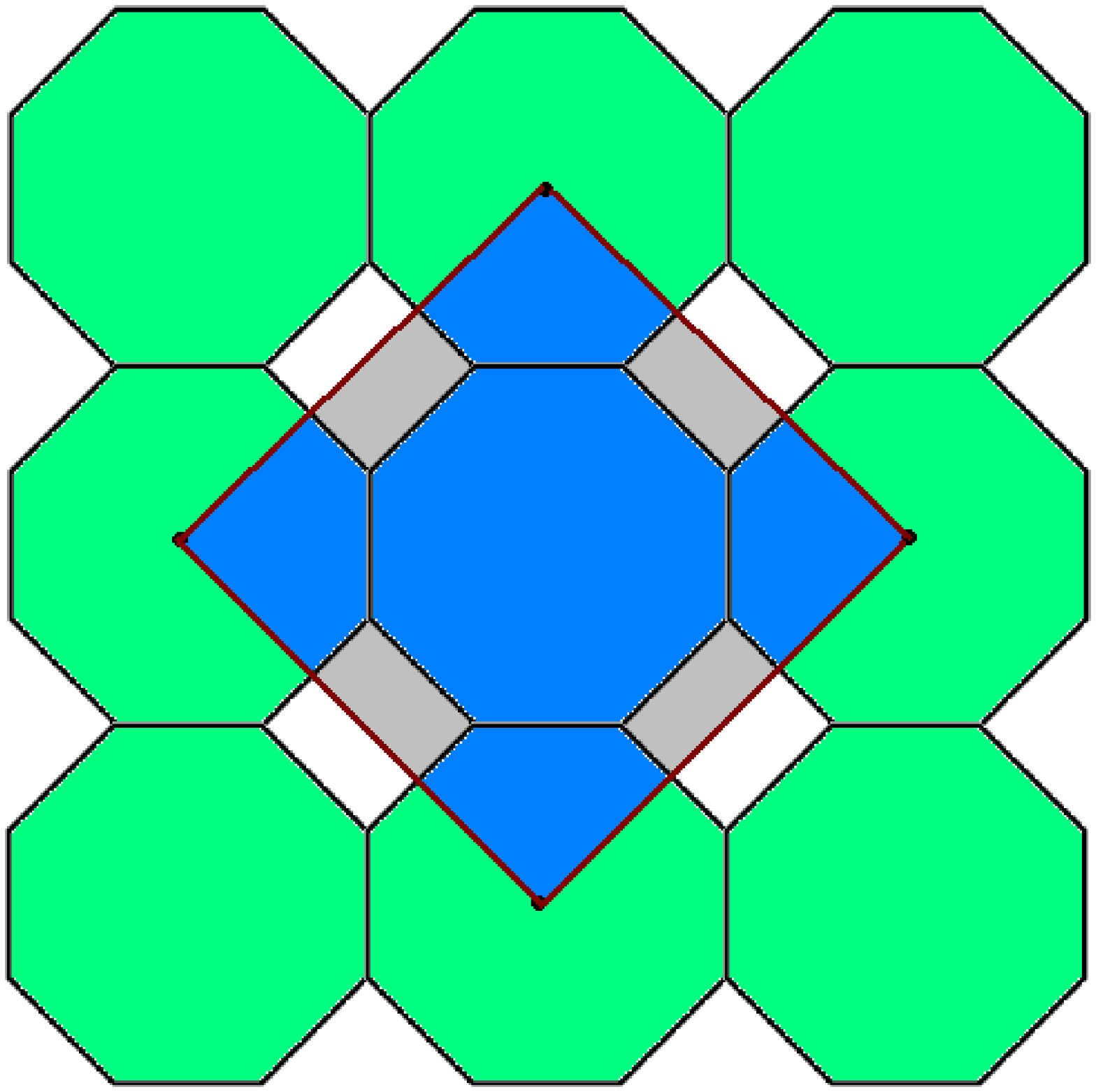} &
\includegraphics[height=3.0cm,keepaspectratio]{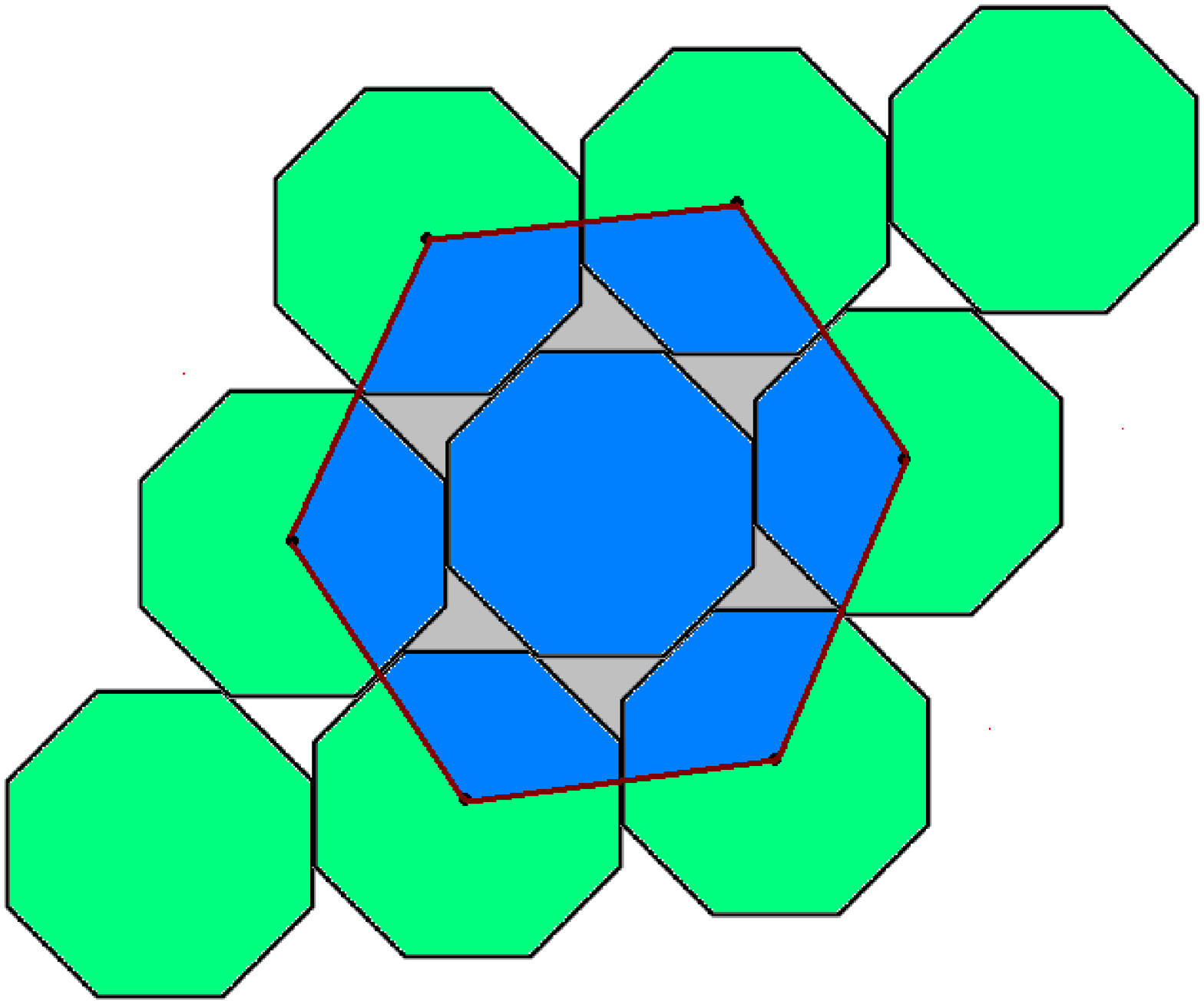} &
\includegraphics[height=3.0cm,keepaspectratio]{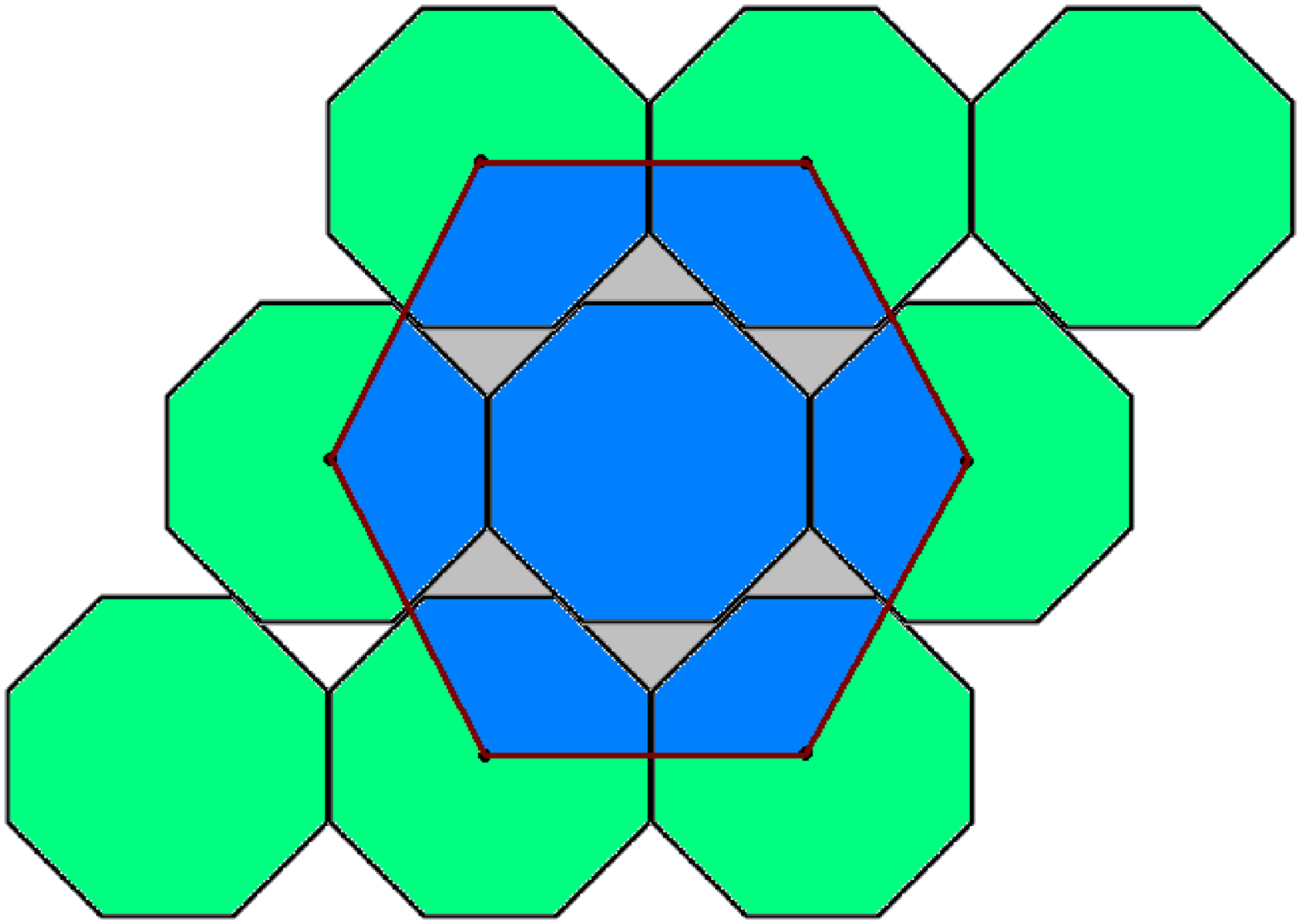} \\
\mbox{\bf (a)} & \mbox{\bf (b)} & \mbox{\bf (c)} & \mbox{\bf (d)}
\end{array}$
\caption{(color online). Illustration of our conjecture
concerning the optimal packings of centrally-symmetric 
particles using two-dimensional octagonal packings.
It is shown how different particle orientations and
arrangements affect the number of face-to-face contacts and
the area of the convex hulls joining the centers of the neighboring 
particles around a central one. (a) The octagons do not have the same
orientation and the number of face-to-face contacts is small.
(b) The octagons are aligned with 4 contacts per particle
to form a lattice packing, but the area of the convex hull associated 
with the central particle is not minimized.
(c) The octagons are aligned with 6 contacts per particle
to form a lattice packing, but although the area of the 
convex hull associated
with the central particle is smaller than in (b),  it is not minimized.
(d) The octagons are aligned with 6 contacts
per particle to form a lattice packing, and the area of 
the convex hull associated with the central particle is minimized.
This minimization corresponds to finding the
the minimal circumscribing hexagon  and therefore corresponds to the
optimal lattice packing \cite{Toth}.}
\label{octagon_pack}
\end{center} 
\end{figure}

 We now sketch what could be the major elements of a proof
of this conjecture. In the case of each of the Platonic solids, 
face-to-face contacts are  favored over vertex-to-face contacts to 
achieve higher packing densities, since the former allows the particle 
centroids to get closer to one another. Such contacting neighbor configurations 
around each particle reduce the volume of the corresponding convex hull joining the
centroids of the contacting particles, and the fraction 
of space covered by the particles within this convex hull should
be increased; see Fig.~\ref{octagon_pack} for a two-dimensional illustration.

To achieve the densest packing, the fraction of space covered by the 
particles within the convex hull should ideally be maximized and so should the
number of face-to-face contacts per particle. Of course, 
this is only a local criterion that may not be consistent 
with the densest global packing.
However, it will be seen that the equivalence of the three principle 
directions of a centrally symmetric
Platonic solid  is crucial for this local 
optimization criterion to be consistent with the globally densest packing.


It is noteworthy that orienting each of the particles in the packings of 
centrally symmetric Platonic solids enable a larger
number of face-to-face contacts and thus allows the maximal fraction 
of space covered by the particles within the convex hull.
For example, a particle with $F$ faces, possesses
$F/2$ families of axes that go through  the centroid of the particle
and intersect the centrally-symmetric face pairs such that
the particles (in the same orientation) with their centroids arranged
on these axes form face-to-face contacts.


The requirement that the particles have the same 
orientation is globally consistent with a lattice packing. 
Indeed, in the optimal lattice packings of the centrally
symmetric Platonic solids, each particle has the maximum number of
face-to-face contacts that could be obtained without violating
the impenetrability condition. It is highly unlikely  
that such particles possessing three equivalent principle 
directions and aligned in the same direction 
could form a more complicated non-lattice periodic packings
with densities that are larger than the optimal lattice packings.
Such non-lattice packings would arise to take advantage of 
the rotational degrees of freedom. By requiring particle alignments, 
only translational degrees of freedom remain
and hence optimization over these degrees of freedom would lead to the globally optimal packings,
which should  be lattice packings. Constraining rotational degrees
of freedom in this way in a non-lattice packing would at best
lead to a local optimum in density.
This conclusion is clearly supported by our simulations, which use 
multiple-particle configurations in the fundamental cell and only produce the optimal lattice 
packings.

In the aforementioned arguments, a key step that is difficult to prove is 
the observation that alignment of each of the particles maximizes
 the number of face-to-face 
contacts and thus the fraction of space covered by the particles within the 
convex hull joining the centroids of the contacting neighbors around a central 
polyhedron. A rigorous verification of this step would lead to the consistency of 
the aforementioned local and global optimization criteria. 


It is noteworthy that in two dimensions, 
our local criterion to achieve the optimal packings 
of centrally symmetric regular polygons (the two-dimensional 
counterparts of the centrally symmetric Platonic solids) 
amounts to the identification of optimal neighbor 
configurations with the maximal edge-to-edge 
contacts per particle such that the convex hull joining the 
centroids of the contacting neighbors is minimal. This statement
is rigorously true in two dimensions because it is tantamount to a theorem due to 
Fejes T{\'o}th \cite{Toth}, which states that the densest packing of a 
centrally symmetric particle in two dimensions can be obtained by 
circumscribing the particle with the minimal centrally symmetric hexagon, which of 
course tiles $\mathbb{R}^2$. Finding the minimal circumscribing
centrally symmetric hexagon is equivalent to finding the minimal area convex hull
joining the centroids of the six contacting particles (the maximal 
number can be obtained in two dimensions for centrally symmetric particles). 
The fact that local
optimality is consistent with the global optimality in two dimensions 
(e.g., a centrally symmetric hexagon can always tessellate the space) 
does not hold in three dimensions. Although we have presented
a three-dimensional generalization of the Fejes T{\'o}th 
theorem, by replacing the minimal hexagon with the minimal convex-hull volume, 
as pointed out before, it is extremely difficult to provide a rigorous proof for it.

Although  Conjecture 1 applies to the centrally symmetric Platonic solids, 
all of our arguments apply as well to each of  the centrally symmetric Archimedean solids.
It cannot be true for the truncated tetrahedron, which is the only non-centrally symmetric 
Archimedean solid. Specifically, it immediately follows from the results of
Ref. \onlinecite{Co06} that a non-lattice packing of truncated tetrahedra can be constructed 
based on the ``primitive Welsh'' tessellation (i.e., by removing the 
small regular tetrahedra) that possesses the 
density $\phi = 23/24 = 0.958333\ldots$, which is appreciably larger than 
the optimal-lattice-packing density of $\phi_{max}^L = 0.680921\ldots$,
and contains two particle centroids per fundamental cell.
In fact, given that the truncated tetrahedra cannot tile space,
the density of the ``primitive Welsh'' packing of truncated tetrahedra 
is so large that it may be the optimal packing of such particles. 
The lattice vectors of this periodic packing are given in Appendix C.

Since the arguments used to justify Conjecture 1 apply equally well
to the 12 centrally symmetric Archimedean solids, we are led to the
following more general conjecture:

\noindent{Conjecture 2:} {\sl The densest packings of the centrally symmetric 
Platonic and Archimedean solids are given by their corresponding optimal lattice packings.}

The aforementioned arguments can also be extended to 
the case of superballs, but here the local principle curvatures 
at the contacting points should be sufficiently small so as to 
maximize the fraction of space covered by the particles
within the convex hull joining the centroids of the neighbors.
The central symmetry and equivalence of the three principle 
directions of superballs means that dense packings
of such objects are favored when the particles are aligned, which
again  leads to  consistency between local and global optimality. The fact that the 
optimal lattice packing densities of superballs is relatively close to 
the upper-bound values, at least around the sphere point, 
as well as results from previous molecular dynamics simulations \cite{Ji09} 
also strongly suggest that the densest packings of these particles 
are given by their corresponding optimal lattice packings.

It is noteworthy that under the assumption that Conjecture 2
is valid, one has upper bounds that are the complete
analog of (\ref{lemma}), i.e.,
\begin{equation}
\phi_{max}\le \mbox{min}\left[\frac{v_p}{v_c}\; \phi_{max}^C,1\right],
\label{lemma2}
\end{equation}
where $v_c$ is the volume of an appropriately chosen centrally
symmetric Platonic or Archimedean solid and $\phi_{max}^C$
is the corresponding optimal density for such a solid, which
according to Conjecture 2 is the optimal lattice packing.
This bound will generally be sharper than bound (\ref{lemma})
because the reference optimal packing is less symmetric
than the sphere. For example, the conjectured bound (\ref{lemma2})
will be sharp for slightly deformed Platonic and Archimedean solids
or nonspherical particles derived by smoothing the vertices, edges
and faces of polyhedra.

\begin{figure}[bthp]
\begin{center} 
$\begin{array}{c@{\hspace{1.0cm}}c}\\
\includegraphics[height=4.5cm,keepaspectratio]{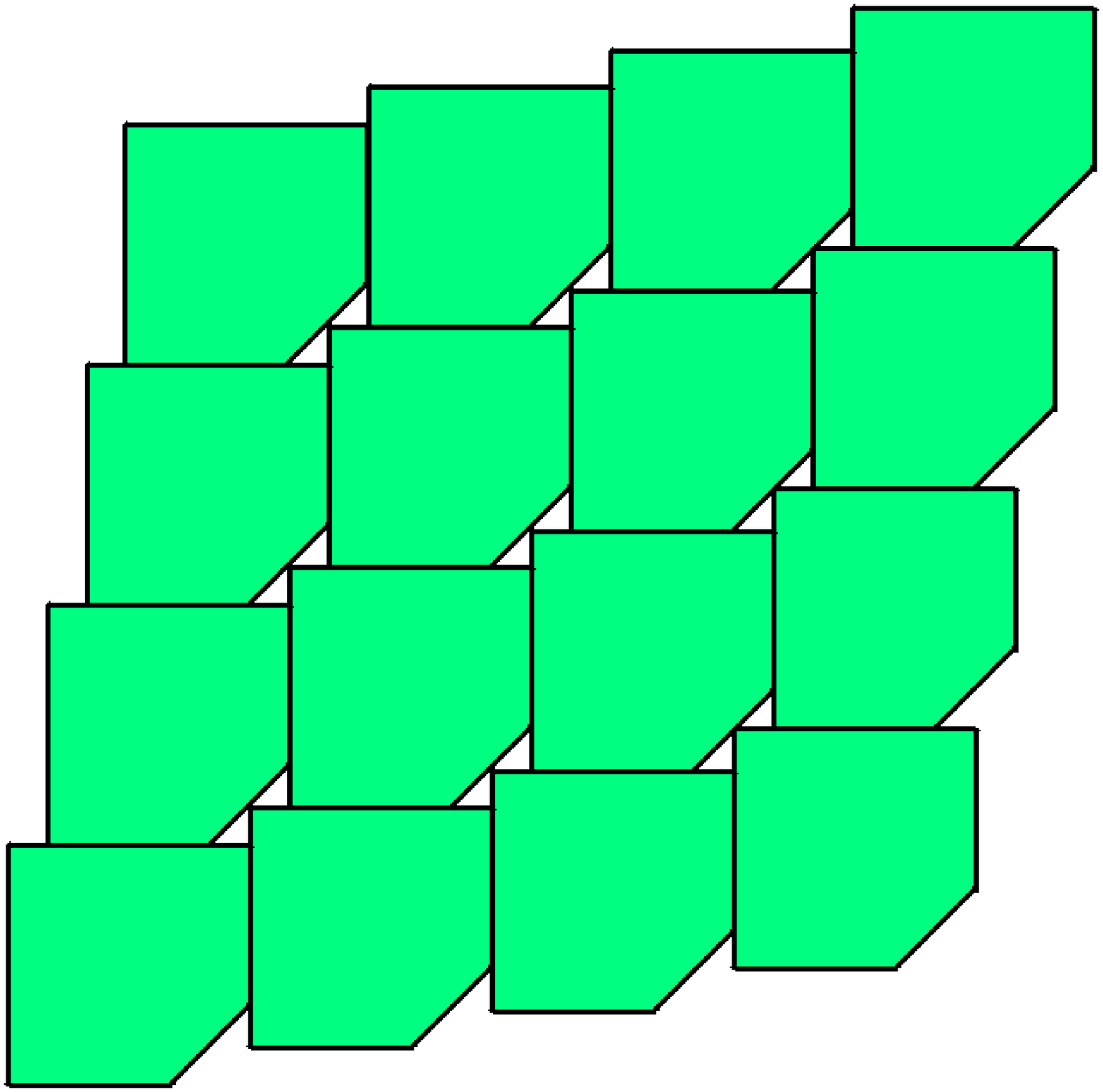} &
\includegraphics[height=4.35cm,keepaspectratio]{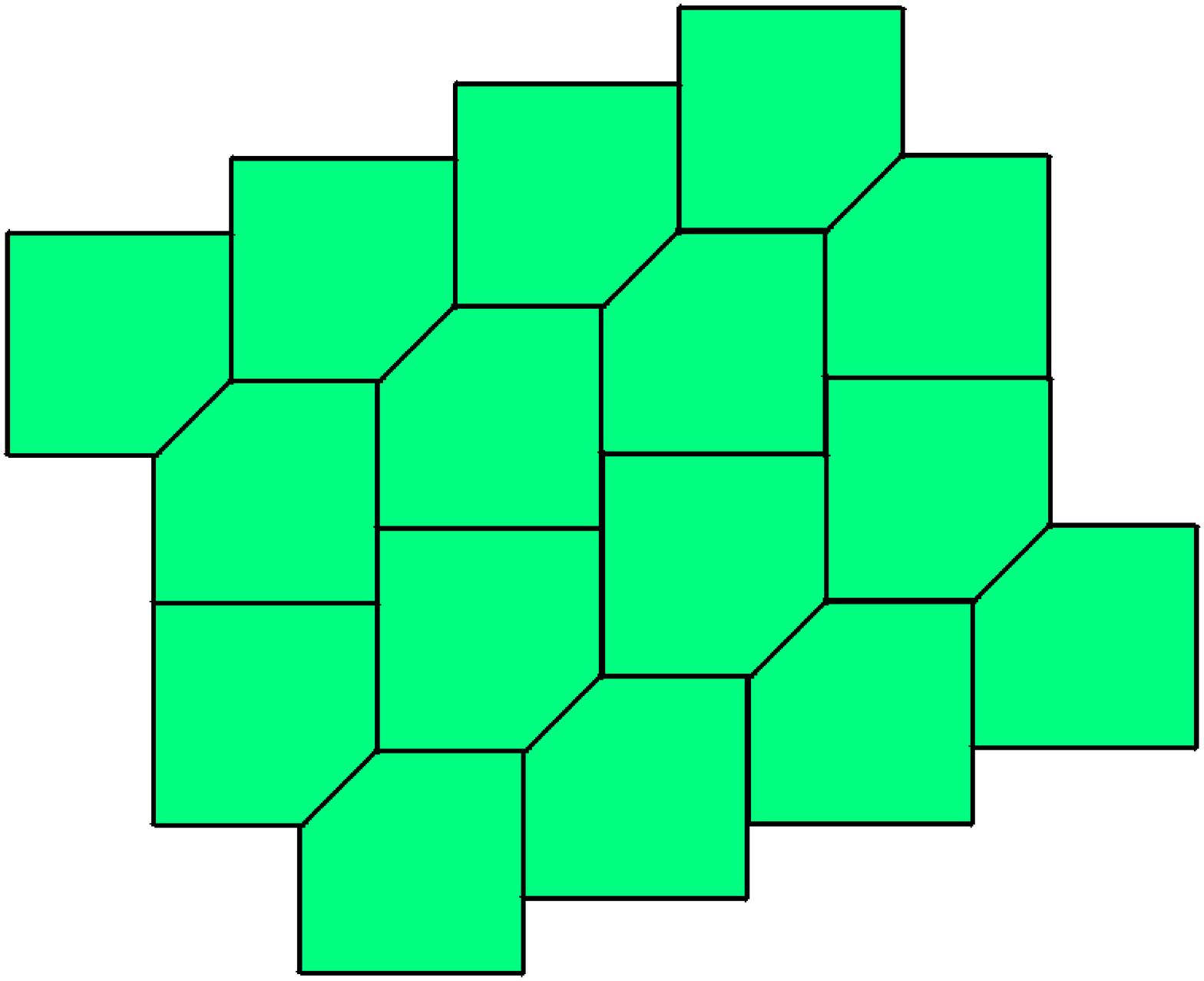} \\
\mbox{\bf (a)} & \mbox{\bf (b)} 
\end{array}$
\caption{(color online). Portions of two packing configurations of pentagons obtained by 
cutting off a corner (isosceles triangle with a right angle) of 
a square. (a) The optimal lattice packing and (b) a two-particle basis periodic packing 
that tiles the plane. Let side length of the square be 1 and the lengths of 
the equal sides of the isosceles triangle be $\beta \in (0,1)$. The lattice 
vectors of the optimal lattice packing are ${\bf e}^L_1 = {\bf i} - \frac{\beta}{2} {\bf j}$, 
${\bf e}^L_2 = (1-\frac{\beta}{2}){\bf i} + (1-\beta) {\bf j}$ and 
the lattice vectors of the periodic packing are ${\bf e}^P_1 = {\bf i} + \beta {\bf j}$,
${\bf e}^P_2 = (1-\beta){\bf i} + (2-\beta) {\bf j}$ with one particle at origin 
and the other at ${\bf b}_1 = (1-\beta){\bf i}+(1-\beta){\bf j}$, where ${\bf i}, {\bf j}$ 
are the unit vectors along the two orthogonal coordinate directions, coinciding 
with two orthogonal sides of the square. The density (covering fraction) of
the optimal lattice packing is $\phi_{max}^L = (1-\frac{\beta^2}{2})/(1-\frac{\beta^2}{4})$ and the 
density of the periodic packing is $\phi_{max}^P = 1$. It can be seen 
that for all $0<\beta<1$, $\phi_{max}^L$ is always smaller than $\phi_{max}^P$.}
\label{Cut_Square}
\end{center} 
\end{figure}

Our work also naturally leads to another conjecture:

\noindent{Conjecture 3:} {\sl The optimal packing of any convex, congruent polyhedron
without central symmetry generally is not a (Bravais) lattice packing.}

\noindent In other words, the set of such polyhedra whose optimal packing 
is not a lattice is overwhelmingly larger than the set whose optimal 
packing is a lattice. We have seen that because the regular tetrahedron and 
truncated tetrahedron lack central symmetry, dense packings
of such objects favor face-to-face contacts. Such orientations
immediately eliminate the possibility that lattice packings (in which particles must
have the same orientations) are optimal. 
 Similarly, it is very plausible that dense packings
of most convex, congruent polyhedra without central symmetry 
are facilitated by face-to-face contacts and hence the optimal packings cannot
be lattices. For example, consider a square with one missing corner, i.e., an 
isosceles triangle with a right angle (see Fig.~\ref{Cut_Square}). 
At first glance, one might surmise that if the missing piece is sufficiently small, the original lattice 
packing should still be optimal or nearly optimal (see Fig.~\ref{Cut_Square}a), 
since lattice packings are optimal for squares. 
However, no matter how small the missing piece may be, a periodic packing 
in which the fundamental cell contains two pentagons can be 
constructed that tile the plane (see Fig.~\ref{Cut_Square}b). 
This is done by taking advantage of the asymmetry of the particle.  
Thus, we see from this counterintuitive example that if 
the particle does not possess central symmetry, it is possible 
to exploit its rotational degrees of freedom to yield
a periodic packing with a complex basis that are generally denser than the 
optimal lattice packing. On the other hand, there are special cases 
where the lattice will be optimal, such as for the rhombic dodecahedron that 
has one corner clipped \cite{hales}. 
However, these special cases are overwhelmed in number by those whose optimal 
packings are not lattices.
If Conjecture 3 is valid, it also applies
to nonspherical particles derived by smoothing the vertices, edges
and faces of polyhedra provided that the particle curvature at 
face-to-face contacts is sufficiently small.

It should not go unnoticed that the densest packings of all
of the Platonic and Archimedean solids reported here as 
well as the densest known packings of superballs \cite{Ji09}
and ellipsoids \cite{Do04} have densities that exceed
the optimal sphere packing density $\phi^S_{max}=\pi/\sqrt{18}=0.7408\ldots$.
These results are consistent with a conjecture of Ulam
who proposed, without any justification, that the optimal
density for packing congruent spheres is smaller than that for any
other convex body \cite{Ulam}. The sphere is perfectly
isotropic with an asphericity $\gamma$ of unity, and therefore,
as noted earlier, its rotational degrees of freedom
are irrelevant in affecting its packing characteristics.
On the other hand, each of the aforementioned convex non-spherical
particles break the continuous rotational symmetry of the sphere
and thus its broken symmetry  can be exploited to yield the densest possible packings.
However, broken rotational symmetry in and of itself
may not be sufficient to satisfy Ulam's conjecture if
the convex particle has a little or no symmetry.

It will also be interesting to determine whether our conjecture 
can be extended to other polyhedral packings. The infinite families of prisms and antiprisms \cite{Cr97} 
provide such a class of packings.  A {\it prism} is a polyhedron having bases that are parallel, 
congruent polygons and sides that are parallelograms. An {\it antiprism} is a polyhedron 
having bases that are parallel, congruent polygons and sides that are alternating bands of triangles.  
Prisms with an even number of sides and antiprisms are centrally symmetric. 
Although prisms and antiprisms are naturally  grouped
with the Archimedean solids (i.e., they are polyhedra in which the same regular polygons appear at
each vertex), they are generally much less symmetric than either the Platonic or Archimedean solids. 
Moreover, even the centrally symmetric prisms and antiprisms generally do not possess three equivalent 
directions. Thus, it is less obvious whether 
Bravais lattices would still provide the optimal packings for these solids, 
except for  prisms that tile space (e.g., hexagonal prism or rhombical prisms). In future work,
it would be desirable to test whether our conjecture extends to prisms
and antiprisms that possess central symmetry and three equivalent directions
using the ASC scheme.

It is worth noting that in four dimensions, the analogs of the tetrahedron,
cube, octahedron, dodecahedron and icosahedron are the four-dimensional 
regular simplex, hypercube, cross polytope, 120-cell
and 600-cell, respectively. All of these four-dimensional polytopes possess central symmetry,
 except for the simplex \cite{Co73}. While the hypercube and cross polytope 
tile $\mathbb{R}^4$,  the optimal packings of simplices are still likely to be non-lattices.
Since our conjecture for the three-dimensional Platonic solids
should still apply in four dimensions, the densest packings of the 
120-cell and 600-cell could be their corresponding optimal lattice packings.
 The cross polytopes for $d \ge 5$
no longer tile space, and their optimal packings may still be their
densest lattice packings provided that $d$ is sufficiently small. However, in 
sufficiently high dimensions, the densest lattice packings of the centrally symmetric polytopes 
are probably no longer optimal, since lattice packings in high dimensions
are known to possess huge ``holes" into which additional particles can be
inserted, yielding higher packing densities with possibly non-lattice
arrangements. Indeed, it has recently been argued that disordered 
sphere packings in very high dimensions could be denser than 
any ordered packing \cite{ToFr06}.

Recent progress in particle synthesis methods have enabled the production of 
a wide spectrum of nanoparticle shapes such as tetrahedra \cite{nano_tetrah}, cubes \cite{nano_cube}, 
icosahedra \cite{nano_icosa} and prisms \cite{nano_prism}. In such applications, it would 
be of great interest to predict the corresponding crystal structures, 
which might possess unusual symmetries and properties. The idea of incorporating collective 
particle motions due to the adaptive fundamental cell should
still make it efficient to search the desired crystal structures formed 
by those polyhedral nano-building blocks. Finally, it is worth noting 
that the ASC algorithm is also suitable to generate 
random packings of polyhedral particles. Crucial dynamical parameters of the system, 
such as the strain rate can be properly controlled to produce packings 
with varying degrees of disorder, including the maximally random jammed ones \cite{Sal00}. 
We will explore disordered packings  in future work.


\section*{ACKNOWLEDGMENTS}
We are grateful to Henry Cohn, John Conway and Alex Jaoshvili
for helpful comments on our manuscript.
S. T. thanks the Institute for Advanced Study for
its hospitality during his stay there.
This work was supported by the Division of Mathematical Sciences
at the National Science Foundation under Award Number DMS-0804431
and by the MRSEC Program of the
National Science Foundation under Award Number DMR-0820341. 
The figures showing the polyhedra were generated 
using the AntiPrism package developed by Adrian Rossiter.

\appendix

\section{Optimal Lattice Packings of the Platonic and Archimedean Solids}

Here we collect fundamental packing characteristics 
(i.e., the lattice vectors, densities and contact numbers) of the optimal 
lattice packings of the Platonic and Archimedean solids, most of 
which are obtained from Refs.~\onlinecite{Mi04}, \onlinecite{Ho70} and \onlinecite{Be00}. 
The Platonic solids are
explicitly defined as the regions (sets of points) bounded by a set of linear
equations of the coordinates. The Archimedean solids are either 
defined as intersections of different Platonic solids or 
delineated based on their symmetry. 
A point set $S$ representing the polyhedron multiplied by 
a number $\alpha$ means an isotropic expansion of the 
polyhedron $S$ with linear ratio $\alpha$. $S_1\cap S_2$ 
represents the intersection region of two polyhedra $S_1$ and $S_2$.
The lattice vectors are given by
column vectors (i.e., the basis vectors are the unit vectors along
the coordinate axis). The naming code used below for the Platonic 
and Archimedean solids is the same one used in Figs.~1 and 3.

A {\it tetrahedron} (P1) has 4 vertices, 6 edges and 4 triangular faces. 
It is defined as the region
\begin{equation}
P1 = \left \{ x\in \mathbb{R}^3: x_1+x_2+x_3\le1,
-x_1-x_2+x_3\le1,  -x_1+x_2-x_3\le1,  x_1-x_2-x_3\le1 \right \}.
\end{equation}
The optimal lattice vectors are given by
\begin{equation}
\begin{array}{c}
{\bf a}_1 = \left ({2, -1/3, -1/3}\right )^T,\quad
{\bf a}_2 = \left ({-1/3, 2, -1/3}\right )^T,\quad
{\bf a}_3 = \left ({-1/3, -1/3, 2}\right )^T,
\end{array}
\end{equation}
where the superscript ``T'' denotes transpose of a column vector.
Each tetrahedron of the packing contacts 14 others.
The packing density is $\phi_{max}^L = 18/49 = 0.367346\ldots$.

An {\it icosahedron} (P2)  has 12 vertices, 30 edges and 20 triangular faces. 
It is defined as the region
\begin{equation}
P2 = \left \{ x\in \mathbb{R}^3: |x_1|+|x_2|+|x_3| \le 1, |\Phi
x_1|+|x_3/\Phi|\le1,  |\Phi x_2|+|x_1/\Phi|\le1,  |\Phi
x_3|+|x_2/\Phi|\le 1 \right \},
\end{equation}
where $\Phi = (1+\sqrt{5})/2$ is the golden ratio. 
The lattice vectors are given by
\begin{equation}
\begin{array}{c}
{\bf a}_1(\bar{x}) = \frac{2}{(1+\Phi)}\left ({\begin{array}{c}
(-33/8-39\sqrt{5}/8)\bar{x}^2+(39/4+33\sqrt{5}/4)\bar{x}-11/4-3\sqrt{5}/2 \\
(-1/4-\sqrt{5}/4)\bar{x}+1+\sqrt{5}/2\\
(33/8+39\sqrt{5}/8)\bar{x}^2+(-19/2-8\sqrt{5})\bar{x}+13/4+3\sqrt{5}/2
                                      \end{array}}\right ),\\\\

{\bf a}_2(\bar{x}) = \frac{2}{(1+\Phi)}\left ({\begin{array}{c}
(-39/8-33\sqrt{5}/40)\bar{x}^2+(35/4+41\sqrt{5}/20)\bar{x}-5/2-23\sqrt{5}/20 \\
(5/4+\sqrt{5}/4)\bar{x}-1-\sqrt{5}/2\\
(-39/8-33\sqrt{5}/40)\bar{x}^2+(15/2+9\sqrt{5}/5)\bar{x}+13/4-3\sqrt{5}/20
                                      \end{array}}\right ),\\\\

{\bf a}_3(\bar(x)) = \frac{2}{(1+\Phi)}\left ({\begin{array}{c}
(3/2+\sqrt{5}/2)\bar{x}-2-\sqrt{5} \\
\bar{x}\\
0
                                      \end{array}}\right ),
\end{array}
\end{equation}
where $\bar{x}\in(1,2)$ is the unique root of the polynomial
\begin{equation}
1086x^3-(1063+113\sqrt{5})x^2+(15\sqrt{5}+43)x+102+44\sqrt{5} = 0.
\end{equation}
It is found that $\bar{x} = 1.59160301\ldots$ and therefore the
lattice vectors, up to nine significant figures,  are given by
\begin{equation}
\begin{array}{c}
{\bf a}_1 = \left ({0.711782425, 0.830400102, 1.07585146}\right )^T,\\
{\bf a}_2 = \left ({-0.871627249, 0.761202911, 0.985203828}\right )^T,\\
{\bf a}_3 = \left ({-0.06919791, 1.59160301, 0}\right )^T.
\end{array}
\end{equation}
Each icosahedron of the packing contacts 12 others.
The packing density is $\phi_{max}^L = 0.836357\ldots$

A {\it dodecahedron} (P3) has 20 vertices, 30 edges and 12 pentagonal faces.
It is defined as the region
\begin{equation}
P3 = \left \{ x\in \mathbb{R}^3: |\Phi x_1|+|x_2|\le1,  |\Phi
x_2|+|x_3|\le1,  |\Phi x_3|+|x_1|\le 1 \right \}.
\end{equation}
The optimal lattice vectors are given by
\begin{equation}
\begin{array}{c}
{\bf a}_1 = \left ({2/(1+\Phi), 2/(1+\Phi), 0}\right )^T,\\
{\bf a}_2 = \left ({2/(1+\Phi), 0, 2/(1+\Phi)}\right )^T,\\
{\bf a}_3 = \left ({0, 2/(1+\Phi), 2/(1+\Phi)}\right )^T.
\end{array}
\end{equation}
Each dodecahedron of the packing contacts 12 others.
The packing density is $\phi_{max}^L = (2+\Phi)/4 = 0.904508\ldots$.

An {\it octahedron} (P4) has 6 vertices, 12 edges and 8 triangular faces.
It is defined as the region
\begin{equation}
P4 = \left \{ x\in\mathbb{R}^3: |x_1|+|x_2|+|x_3|\le1 \right \}.
\end{equation}
The optimal lattice vectors are given by
\begin{equation}
\begin{array}{c}
{\bf a}_1 = \left ({2/3, 1, 1/3}\right )^T,\quad
{\bf a}_2 = \left ({-1/3, -2/3, 1}\right )^T,\quad
{\bf a}_3 = \left ({-1, 1/3, -2/3}\right )^T.
\end{array}
\end{equation}
Each octahedron of the packing contacts 14 others.
The packing density is $\phi_{max}^L = 18/19 = 0.947368\ldots$.

A {\it cube} (P5)  has 8 vertices, 12 edges and 6 square faces.
It is defined as the region
\begin{equation}
P5 = \left \{ x\in\mathbb{R}^3: |x_i|\le1 \right \}.
\end{equation}
The optimal lattice vectors are given by
\begin{equation}
\begin{array}{c}
{\bf a}_1 = \left ({2, 0, 0}\right )^T,\quad
{\bf a}_2 = \left ({0, 2, 0}\right )^T,\quad
{\bf a}_3 = \left ({0, 0, 2}\right )^T.
\end{array}
\end{equation}
Each cube of the packing contacts 26 others (which includes
vertex-to-vertex contacts).
The packing density is $\phi_{max}^L = 1$.

A {\it truncated tetrahedron} (A1) has 12 vertices, 18 edges 
and 8 face: 4 hexagons and 4 triangles.
It is defined as the region
\begin{equation}
A1 = \left \{ x\in\mathbb{R}^3: x \in 5\cdot P1 \cap -3\cdot P1 \right\}.
\end{equation}
The optimal lattice vectors are given by
\begin{equation}
\begin{array}{c}
{\bf a}_1 = \left ({4/3, 4, 8/3}\right )^T,\quad
{\bf a}_2 = \left ({4, -8/3, -4/3}\right )^T,\quad
{\bf a}_3 = \left ({-8/3, 4/3, -4}\right )^T.
\end{array}
\end{equation}
Each truncated tetrahedron of the packing contacts 14 others. 
The packing density is $\phi_{max}^L = 207/304 = 0.680921\ldots$.

A {\it truncated icosahedron} (A2) has 60 vertices, 90 edges 
and 32 faces: 20 hexagons and 12 pentagons. 
It is defined as the region 
\begin{equation}
A2 =  \left \{ x\in\mathbb{R}^3: x \in (1+\Phi)\cdot P2 \cap (4/3+\Phi)\cdot P3 \right\}.
\end{equation}
The optimal lattice vectors are given by
\begin{equation}
\begin{array}{c}
{\bf a}_1 = (1+\Phi){\bf a}^{P2}_1,\quad
{\bf a}_2 = (1+\Phi){\bf a}^{P2}_2,\quad
{\bf a}_3 = (1+\Phi){\bf a}^{P2}_3,
\end{array}
\end{equation}
where ${\bf a}^{P2}_i$ ($i=1,2,3$) are the lattice vectors 
of the optimal lattice packing of icosahedra.
Each truncated icosahedron of the packing contacts 12 others.
The packing density is $\phi_{max}^L = 0.784987\ldots $

A {\it snub cube} (A3) has 24 vertices, 60 edges 
and 38 faces: 6 squares and 32 triangles.
Let the snub cube orientated in a way such that its 
6 square faces lie in the hyperplanes 
\begin{equation}
\left \{ x\in\mathbb{R}^3: |x_i|=1 (i=1,2,3)\right \}.
\end{equation}
The optimal lattice vectors are given by
\begin{equation}
\begin{array}{c}
{\bf a}_1 = \left ({2, 0, 0}\right )^T,\quad
{\bf a}_2 = \left ({0, 0, 2}\right )^T,\quad
{\bf a}_3 = \left ({1, 2/y^*-2, -1}\right )^T.
\end{array}
\end{equation}
where $y^*$ is the unique real solution of $y^3 + y^2 + y = 1$.
Each snub cube of the packing contacts 12 others.
The packing density is $\phi_{max}^L = 0.787699\ldots $

A {\it snub dodecahedron} (A4) has 60 vertices, 150 edges 
and 92 faces: 12 pentagons and 80 triangles.
Let the snub dodecahedron orientated in a way such that its 
12 pentagonal faces lie in the hyperplanes of the faces 
of the dodecahedron $(1+\Phi)\cdot P3$. 
The optimal lattice vectors are given by
\begin{equation}
\begin{array}{c}
{\bf a}_1 = \left ({2, 2, 0}\right )^T,\quad
{\bf a}_2 = \left ({2, 0, 2}\right )^T,\quad
{\bf a}_3 = \left ({0, 2, 2}\right )^T.
\end{array}
\end{equation}
Each snub dodecahedron of the packing contacts 12 others.
The packing density is $\phi_{max}^L = 0.788640\ldots $

A {\it rhombicosidodecahdron} (A5) (also known as small rhombicosidodecahedron) 
has 60 vertices, 120 edges and 62 face: 12 pentagons, 30 squares and 20 triangles.
It is defined as the region:
\begin{equation}
A5 = \left \{ x\in\mathbb{R}^3: x \in (3\Phi+2)\cdot [P5\cap(1+\sqrt{2})P4] \cap (4\Phi+1)\cdot P2 
\cap (3+3\Phi)\cdot P3 \right\}.
\end{equation}
The optimal lattice vectors are given by
\begin{equation}
\begin{array}{c}
{\bf a}_1 = \left ({(\Phi-1)/(2\Phi+1), 7, (9\Phi+4)/(2\Phi+1)}\right )^T,\\
{\bf a}_2 = \left ({(9\Phi+4)/(2\Phi+1), (\Phi-1)/(2\Phi+1), 7}\right )^T,\\
{\bf a}_3 = \left ({7, (9\Phi+4)/(2\Phi+1), (\Phi-1)/(2\Phi+1)}\right )^T.
\end{array}
\end{equation}
Each rhombicosidodecahedron of the packing contacts 12 others. 
The packing density is $\phi_{max}^L = (8\Phi+46)/(36\Phi+15) = 0.804708\ldots$.

A {\it truncated icosidodecahdron} (A6) (also known as great rhombicosidodecahedron) 
has 120 vertices, 180 edges and 62 faces: 12 decagons, 20 hexagons and 30 squares. 
It is defined as the region
\begin{equation}
A6 = \left \{ x\in\mathbb{R}^3: x \in (5\Phi+4)\cdot [P5\cap(1+\sqrt{2})P4] \cap (6\Phi+3)\cdot P2 
\cap (5+5\Phi)\cdot P3 \right\}.
\end{equation}
The optimal lattice vectors are given by
\begin{equation}
\begin{array}{c}
{\bf a}_1 = \left ({10, 10, 0}\right )^T,\quad
{\bf a}_2 = \left ({10, 0, 10}\right )^T,\quad
{\bf a}_3 = \left ({0, 10, 10}\right )^T.
\end{array}
\end{equation}
Each truncated icosidodecahedron  of the packing contacts 12 others. 
The packing density is $\phi_{max}^L = (2\Phi/5+9/50) = 0.827213\ldots$.

A {\it truncated cubeoctahedron} (A7) (also known as great rhombicubeoctahedron) 
has 48 vertices, 72 edges and 26 faces: 6 octagons, 8 hexagons and 12 squares.
It is defined as the region
\begin{equation}
\begin{array}{c}
A7 = \{ {x\in\mathbb{R}^3: |x_1|+|x_2|\le (2+3\sqrt{2}), |x_2|+|x_3|\le (2+3\sqrt{2}),} \\
   \quad\quad {|x_2|+|x_3|\le (2+3 \sqrt{2}), x \in (2 \sqrt{2}+1)\cdot P5 \cap (3\sqrt{2}+3)\cdot P4 } \}.
\end{array}
\end{equation}
The optimal lattice vectors are given by
\begin{equation}
\begin{array}{c}
{\bf a}_1 = \left ({4\sqrt{2}+2, -4\sqrt{2}-1+2\alpha, 4\sqrt{2}+1-2\alpha}\right )^T,\\
{\bf a}_2 = \left ({\sqrt{2}/2-3/2+\alpha, -3\sqrt{2}/2+1/2+\alpha, 4\sqrt{2}+2}\right )^T,\\
{\bf a}_3 = \left ({7/2+7\sqrt{2}/2-\alpha, 1+2\alpha, 5\sqrt{2}/2+3/2-\alpha}\right )^T,
\end{array}
\end{equation}
where $\alpha = \sqrt{33}(\sqrt{2}+1)/6$.
Each truncated cubeoctahedron of the packing contacts 12 others. 
The packing density is $\phi_{max}^L = 0.8493732\ldots$.

An {\it icosidodecahedron} (A8) has 30 vertices, 60 edges 
and 32 faces: 12 pentagons and 20 triangles. 
It is defined as the region
\begin{equation}
A8 = \left \{ x\in\mathbb{R}^3: x \in  P2\cap P3 \right\}.
\end{equation}
The optimal lattice vectors are given by
\begin{equation}
\begin{array}{c}
{\bf a}_1 = \left ({2/(1+\Phi), 2/(1+\Phi), 0}\right )^T,\\
{\bf a}_2 = \left ({2/(1+\Phi), 0, 2/(1+\Phi)}\right )^T,\\
{\bf a}_3 = \left ({0, 2/(1+\Phi), 2/(1+\Phi)}\right )^T.
\end{array}
\end{equation}
Each icosidodecahedron of the packing contacts 12 others. 
The packing density is $\phi_{max}^L = (14+17\Phi)/48 = 0.864720\ldots$.

A {\it rhombicuboctahedron} (A9) (also known as small rhombicubeoctahedron) 
has 24 vertices, 48 edges and 26 faces: 18 squares and 8 triangles. 
It is defined as the region 
\begin{equation}
\begin{array}{c}
A9 = \{ {x\in\mathbb{R}^3: |x_1|+|x_2|\le 2, |x_2|+|x_3|\le 2,} \\
   \quad\quad  {|x_1|+|x_3|\le 2,x \in \sqrt{2}\cdot P5 \cap (4-\sqrt{2})\cdot P4} \}.
\end{array}
\end{equation}
The optimal lattice vectors are given by
\begin{equation}
\begin{array}{c}
{\bf a}_1 = \left ({2, 2, 0}\right )^T,\quad
{\bf a}_2 = \left ({2, 0, 2}\right )^T,\quad
{\bf a}_3 = \left ({0, 2, 2}\right )^T.
\end{array}
\end{equation}
Each truncated cubeoctahedron of the packing contacts 12 others. 
The packing density is $\phi_{max}^L = (16\sqrt{2}-20)/3 = 0.875805\ldots$.

A {\it truncated dodecahedron} (A10) has 60 vertices, 90 edges 
and 32 faces: 12 decagons and 20 triangles.
It is defined as the region 
\begin{equation}
\begin{array}{c}
A10 = \left \{ x \in (1+\Phi)\cdot P3 \cap [(7+12\Phi)/(3+4\Phi)]\cdot P2 \right\}.
\end{array}
\end{equation}
The optimal lattice vectors are given by
\begin{equation}
\begin{array}{c}
{\bf a}_1 = \left ({2, 2, 0}\right )^T,\quad
{\bf a}_2 = \left ({2, 0, 2}\right )^T,\quad
{\bf a}_3 = \left ({0, 2, 2}\right )^T.
\end{array}
\end{equation}
Each truncated dodecahedron of the packing contacts 12 others. 
The packing density is $\phi_{max}^L = (5\Phi+16)/(24\Phi-12) = 0.897787\ldots$.

A {\it cuboctahedron} (A11) has 12 vertices, 24 edges 
and 14 faces: 6 squares and 8 triangles. 
It is defined as the region
\begin{equation}
\begin{array}{c}
A11 = \left \{ x \in P5 \cap 2\cdot P4 \right\}.
\end{array}
\end{equation}
The optimal lattice vectors are given by
\begin{equation}
\begin{array}{c}
{\bf a}_1 = \left ({2, -1/3, -1/3}\right )^T,\quad
{\bf a}_2 = \left ({-1/3, 2, -1/3}\right )^T,\quad
{\bf a}_3 = \left ({-1/3, -1/3, 2}\right )^T.
\end{array}
\end{equation}
Each cuboctahedron of the packing contacts 14 others. 
The packing density is $\phi_{max}^L = 45/49 = 0.918367\ldots$.

A {\it truncated cube} (A12) has 24 vertices, 36 edges 
and 14 faces: 6 octagons and 8 triangles.
It is defined as the region
\begin{equation}
\begin{array}{c}
A12 = \left \{ x \in P5 \cap (1+\sqrt{2})\cdot P4 \right\}.
\end{array}
\end{equation}
The optimal lattice vectors are given by
\begin{equation}
\begin{array}{c}
{\bf a}_1 = \left ({2, -2\alpha, 0}\right )^T,\quad
{\bf a}_2 = \left ({0, 2, -2\alpha}\right )^T,\quad
{\bf a}_3 = \left ({-2\alpha, 0, 2}\right )^T,
\end{array}
\end{equation}
where $\alpha = (2-\sqrt{2})/3$.
Each truncated cube of the packing contacts 14 others. 
The packing density is $\phi_{max}^L = 9/(5+3\sqrt{2}) = 0.973747\ldots$.

A {\it truncated octahedron} (A13) has 24 vertices, 36 edges 
and 14 faces: 8 hexagons and 6 squares.
It is defined as the region
\begin{equation}
\begin{array}{c}
A13 = \left \{ x \in P5 \cap (3/2)\cdot P4 \right\}.
\end{array}
\end{equation}
The optimal lattice vectors are given by
\begin{equation}
\begin{array}{c}
{\bf a}_1 = \left ({2, 0, 0}\right )^T,\quad
{\bf a}_2 = \left ({2, 2, 0}\right )^T,\quad
{\bf a}_3 = \left ({1, 1, -1}\right )^T.
\end{array}
\end{equation}
Each truncated octahedron of the packing contacts 14 others. 
The packing density is $\phi_{max}^L = 1$.

\section{Lattice Vectors and Other Characteristics of the Densest Known Tetrahedral Packing}

Here we report the lattice vectors of fundamental cell for the densest known 
tetrahedral packing with $314$ particles up to 12 significant figures, 
although our numerical precision is not limited by that. The side length of the tetrahedron 
is $d_0 = 1.8$ and its volume is $v_p = \sqrt{2}  d_0^3/12 =  0.687307791313$.
The lattice vectors are given by
\begin{equation}
\begin{array}{c}
{\bf a}_1 = (6.130348985438,~-1.714011427194,~0.022462185673)^T, \\
{\bf a}_2 = (1.622187720300,~6.106123418234,~0.123567805838)^T, \\
{\bf a}_3 = (0.055014355972,~0.116436714923,~6.526427521512)^T.
\end{array}
\end{equation}
The volume of the fundamental cell is
$\mbox{Vol}(F) = |{\bf a}_1\times{\bf a}_2\cdot{\bf a}_3| = 262.344828467328$.
Thus, the packing density is readily computed as
\begin{equation}
\phi = \frac{N \cdot v_p}{\mbox{Vol}(F)} =
\frac{314\times 0.687307791313}{262.344828467328} = 0.822637319490.
\end{equation}
The coordinates of each of the 314 tetrahedra are given elsewhere \cite{web}.


\section{Lattice Vectors of the Densest Known Packing of Truncated Tetrahedra}

The ``primitive Welsh'' tessellation of space consists of truncated large 
regular tetrahedra and small regular tetrahedra, as described in Ref.~\onlinecite{Co06}.
When the small tetrahedra are removed, the remaining truncated tetrahedra
 give the densest known periodic (non-lattice) packing of 
such objects. Following Ref.~\onlinecite{Co06}, the centroids 
of the truncated tetrahedra sit at the nodes {\bf 0} and {\bf 1} of 
the two adjacent cells of a body-centered cubic lattice shown in Fig.~2(a) of 
Ref.~\onlinecite{Co06}. Let length of the edges of the truncated tetrahedron 
be $d_E = \sqrt{2}/2$, then the lattice vectors are given by 
\begin{equation}
\begin{array}{c}
{\bf a}_1 = (1, 1, 0)^T, \quad
{\bf a}_2 = (1, 0, 1)^T, \quad
{\bf a}_3 = (0, 1, 1)^T.
\end{array}
\end{equation}
The basis vectors for the centroids of {\bf 0}-type and {\bf 1}-type 
truncated tetrahedra are respectively given by 
\begin{equation}
\begin{array}{c}
{\bf b}_0 = (1/2, 1/2, 1/2)^T, \quad {\bf b}_1 = (0, 0, 0)^T. 
\end{array}
\end{equation}


\begin{thebibliography}{30}
\bibitem{overlap}
The overlap potential function of a pair of strictly convex and
smooth particles is an analytic function of the positions, orientations and
shapes of the two particles, whose value indicates whether the two particles 
overlap or not, or whether they are tangent to one another; see also Ref. \onlinecite{Do05a}.
\bibitem{Deform}
In this paper, a deformation refers to a volume-preserving shape change of the fundamental cell
(i.e., a shear), which is to be distinguished from a volume change
of the cell, which can either be a compression or an expansion of the cell.
\bibitem{conjecture}
In Ref. \onlinecite{Co06}, the authors suggested (not conjectured) that
the regular tetrahedron ``may not be able to pack as densely as
the sphere, which would contradict a conjecture of Ulam." Indeed,
the results of Ref. \onlinecite{Ch08} verifies that the tetrahedron
does not violate Ulam's conjecture.
\bibitem{footnoteTetrah}
The fundamental cells of the initial configurations for the optimization 
algorithm involve periodic copies of the packings listed in Table~I. 
Specifically, if the lattice vectors of the packings listed in Table~I 
are denoted by ${\bf e}_1$, ${\bf e}_2$ and ${\bf e}_3$, then the lattice vectors 
associated with the initial configurations are given 
by ${\bf e}^L_1 = n_1 {\bf e}_1$,  ${\bf e}^L_2 = n_2 {\bf e}_2$ and 
${\bf e}^L_3 = n_3 {\bf e}_3$, where $n_1, n_2, n_3 \ge 1$ are integers. 
Thus, the number of tetrahedra in the 
fundamental cells of the initial configurations is an integral multiple, i.e., 
$n_1\times n_2 \times n_3$, of that for the corresponding packings given in Table~I. 
For the optimal lattice packing, $n_1 = n_2 = n_3 = 3$; for the uniform
packing, $n_1 = n_2 = n_3 = 3$; for the Welsh packing, $n_1 = n_2 = n_3 =1$; 
for the icosahedral packing, $n_1 = n_2 = n_3 = 1$; and for the wagon-wheels 
packing, $n_1 = n_2 = 2, n_3 = 1$.
\bibitem{footnoteBarlow}
Although we cannot  exclude the possibility that there exist other 
optimal packing particle arrangements that possess the same density 
as the optimal lattice packings, it is highly unlikely 
that those arrangements could be constructed in the same way
way as the stacking variants  of the optimal FCC lattice packings
for spheres (the so-called Barlow packings). The polyhedral shape of the particles 
does not allow them  to fit snugly into the ``holes'' formed in the particle layers
above and below a particular layer, 
as in the case of spheres, and hence stackings of the layers 
cannot result in packings with densities that are higher than that of the optimal lattice packing. 
If degenerate optimal packings exist, they cannot be of the ``Barlow" type.
\end{thebibliography}
\end{document}